\documentclass[english]{article}
\usepackage{geometry}
\usepackage{amsmath}
\usepackage{graphicx}
\usepackage{amssymb}
\usepackage[authoryear]{natbib}

\makeatletter
\newcommand{\lyxrightaddress}[1]{
\par {\raggedleft \begin{tabular}{l}\ignorespaces
#1
\end{tabular}
\vspace{1.4em}
\par}
}

\usepackage{a4wide}
\usepackage{graphicx}
\usepackage{amssymb}

\newcommand{\mathe}{\mathrm{e}}
\newcommand{\mathd}{\mathrm{d}}

\usepackage{babel}

\usepackage{babel}
\makeatother
\begin{document}

\title{The mean electro-motive force and current helicity under the influence
of rotation, magnetic field and shear. }

\author{Pipin V.V.%
\thanks{pip@iszf.irk.ru%
}}

\maketitle

\lyxrightaddress{Institute Solar-Terrestrial Physics, Irkutsk, Russia.}

\begin{abstract}
The  mean electromotive force (MEMF) in a rotating stratified 
magnetohydrodynamical turbulence is studied.
Our study rests on the  mean-field
magnetohydrodynamics framework and $\tau$ approximation. We compute the effects of the large-scale magnetic
fields (LSMF),  global rotation and large-scale shear flow on the
different parts of the MEMF (such
as $\alpha$ - effect, turbulent diffusion, turbulent transport, etc.)
in an explicit form. The influence of the helical magnetic fluctuations which stem
from the small-scale dynamo is taken into
account, as well. In the paper,  we derive the equation
governing the current helicity evolution. It is shown that the joint
effect of the differential rotation and magnetic fluctuations in the
stratified media can be responsible for the generation, maintenance
and redistribution of the current helicity. The implication of the
obtained results to astrophysical dynamos is considered as
well. 
\end{abstract}

\section{Introduction}

The mean-field magnetohydrodynamics presents one of the most powerful
tools for exploring the nature of the large-scale magnetic activity
in cosmic bodies \citep{moff:78,park,krarad80}. It is widely believed
that magnetic field generation there is governed by interplay between
turbulent motions of electrically conductive fluids and global rotation.
The growth and evolution of the large-scale magnetic fields (LSMF)
in cosmic plasma strongly depends on the mean electromotive force,
$\mathbf{\mathcal{E}}=\left\langle {\mathbf{u}\times\mathbf{b}}\right\rangle $,
which is given by the correlation between the fluctuating components
of the velocity field of plasma, $\mathbf{u}$, and the fluctuating
magnetic fields, $\mathbf{b}$.

The global rotation, stratification and the strong LSMF can substantially
modify the structure and amplitude of the mean electromotive force
(hereafter, MEMF) leading to the rich variety of the turbulence effects
driving the evolution of the LSMF in cosmic bodies, e.g., the $\alpha$-effect
\citep{rob-saw,moff:78,krarad80,park,kit-rud:1993b}, the rotationally-induced
anisotropy of turbulent diffusion and effective drift of LSMF \citep{rob-saw,krarad80,kit-pip-rud},
etc.. Generally speaking, the nonlinear effects of the small-scale
Lorentz forces on the MEMF and LSMF evolution stem from two sources.
One is driven by perturbations of the LSMF due to turbulent motions
and another one is due to magnetic fluctuations, which are maintained
by the small-scale dynamo action in a turbulent medium. The role of
the small-scale dynamo in the LSMF evolution is still insufficiently
understood. Numerous contributions to this subject can be found in
the modern literature, e.g., \citep{moff:78,pouquet-al:1975a,pouquet-al:1975b,bra-sub:04}.
According to the mentioned studies the most important effect of the
growing magnetic fluctuations on the LSMF evolution is caused by the
helical part of magnetic fluctuations. The magnetic helicity conservation
law, if applied to the mean-field magnetohydrodynamics, requires that
the amount of helicity contained in the LSMF (controlled mostly by
$\alpha$-effect) should be roughly the same and opposite in sign
to its counterpart in the small scales, see \citep{kleruz82}. In
this way the helical part of magnetic fluctuations, which is excited
both due to shredding the LSMF by turbulent motions and due to small-scale
dynamo, effectively saturates the generation of the LSMF by $\alpha$-effect
(\citet{vain-kit:83,bran:01,bf:02a,black-bran:02}). Further discussions
on this subject can be found in above cited papers. Their main lesson
is that the construction of the realistic mean-field dynamo theory
requires the evolution of the small scale magnetic (or current-) helicity
to be taken into account.

Currently, there are two basic schemes for computing the MEMF of turbulent
fields. One is the quasi-linear approximation (the same approximation
is called the FOSA or SOCA in literature). A comprehensive discussion
about its applicability and validity in astrophysics can be found
at papers by \citet{moff:78,park,krarad80,bra-sub:04}. This scheme
remains one of the main tools of the mean-field magnetohydrodynamics.
However, one of unfortunate problem of SOCA is that the contribution
of the magnetic fluctuations (and the corresponding magnetic helicity)
driven by the small-scale dynamo is hardly possible to include in
the theory in self-consistent way. The third order closure scheme
based on $\tau$-approximation (\citet{ors:70,vain-kit:83,rad-kle-rog,bra-sub:04})
gives a chance to consider, roughly, the effects of the small-scale
dynamo on the MEMF. Following \citep{bra-sub:04} (hereafter BS05),
I will call it MTA (minimal tau approximation). Different kinds of
this approximation are used in the literature, see \citep{vain:83,vain-kit:83,rad-kle-rog,bra-sub:04,kle-rog:04a,bf:02,bf:02a}.
In the paper we follow procedure described in BS05. Furthermore, the
variant of tau approximation with a scale-independent relaxation time,
$\tau$, is applied. For this reason, some results obtained in the
paper can be different of those that are given elsewhere: \citep{kle-rog:04a,kle-rog:04b,kle-rog:04c,rad-kle-rog}. 

The main purpose of this paper is to compute the MEMF via MTA taking
into account the influence of the global rotation and LSMF on the
turbulence. The stratification of the medium and the large-scale shear
are taken into account as well. The influence of rotation, LSMF and
uniform shear on the different parts of the MEMF (such as $\alpha$
- effect, turbulent diffusion, turbulent transport and etc.) is explicitly
defined via factors describing the efficiency of rotational and LSMF
feedback on the turbulent flows. The influence of rotation is measured
by the Coriolis number, $\Omega^{\ast}=2\Omega\tau_{c}$ , where $\Omega$
is the solid body angular velocity and $\tau_{c}$ - the typical correlation
time of turbulent flows. The influence of LSMF is measured by $\beta=\bar{B}/\left(u_{c}\sqrt{\mu\rho}\right)$,
where $\bar{B}$ is the strength of the LSMF, $u_{c}$ is a typical
rms velocity of turbulent flows and $\mu$, $\rho$ are the magnetic
permeability and the density of the media, respectively. Following
the basic approach developed in above cited papers we derive the equations
governing the evolution of the current helicity both in rotating and
in magnetized turbulent flows with imposed uniform shear.

The paper is structured as follows. In the next section we shortly
outline the basic equations, assumptions and the computational scheme
for derivation of the MEMF and the evolutionary equation for current
helicity. Section 3 is devoted to the results of calculations of the
MEMF for different situations (slow rotation, strong LSMF, vice versa
and etc.). In section 4 we derive the evolutionary equation for current
helicity. In section 5 we summarize the main results of the paper.

\section{Basic equations}
In the spirit of the mean-field magnetohydrodynamics, we split the
physical quantities of the turbulent conducting fluid into the mean
and randomly fluctuating part with the mean part defined as the ensemble
average of the random fields. One assumes the validity of the Reynolds
rules. The magnetic field $\mathbf{B}$ and velocity of motions $\mathbf{V}$
are decomposed as follows: $\mathbf{B}=\overline{\mathbf{B}}+\mathbf{b}$,
$\mathbf{V}=\overline{\mathbf{V}}+\mathbf{u}$. Hereafter, everywhere,
we use the small letters for the fluctuating part of the fields and
capital letters with a bar above for the mean fields. The angle brackets
are used for the ensemble average of products. Following the lines
of two-scale approximation \citep{rob-saw,krarad80} we assume that
the mean fields vary over the much larger scales (both in time and
in space) than the fluctuating fields. The average effect of the MHD-turbulence
on the LSMF evolution is described by the MEMF, $\mathbf{\mathcal{E}}=\left\langle {\mathbf{u\times b}}\right\rangle $.
The governing equations for fluctuating magnetic field and velocity
are written in a rotating coordinate system as follows
 \begin{eqnarray}
\frac{\partial\mathbf{b}}{\partial t} & = & \nabla\times\left(\mathbf{u}\times\mathbf{\overline{B}}+\mathbf{\overline{V}}\times\mathbf{b}\right)+\eta\nabla^{2}\mathbf{b}+{\mathfrak{G}},\label{induc1}\\
\frac{\partial m_{i}}{\partial
  t}+2\left(\mathbf{\Omega}\times\mathbf{m}\right)_{i} & = & 
-\nabla_{i}\left(p-\frac{2}{3}\left(\mathbf{G\cdot m}\right)\nu+\frac{\left(\mathbf{b\cdot\overline{B}}\right)}{2\mu}\right)+\nu\Delta m_{i}+\nu\left(\mathbf{G\cdot\nabla}\right)m_{i}\label{navie1}\\
 & + &
 \frac{1}{\mu}\nabla_{j}\left(\overline{B}_{j}b_{i}+\overline{B}_{i}b_{j}\right)
-\nabla_{j}\left(\overline{V}_{j}m_{i}+\overline{V}_{i}m_{j}\right)+f_{i}+\mathfrak{F}_{i},\nonumber \end{eqnarray}
where $\mathfrak{G},\mathfrak{F}$ are nonlinear contributions
of fluctuating fields, ${\mathbf{m}}=\bar{\rho}{\mathbf{u}}$,
$\mathbf{G}=\nabla\log\bar{\rho}$ is the density stratification scale
of the media, $p$ - the fluctuating pressure, $\mathbf{\Omega}$
- the angular velocity responsible for the Coriolis force, $\mathbf{\overline{V}}$
- mean flow which is a weakly variable in space, $\mathbf{f}$ - the
random force driving the turbulence. 

To compute $\mathbf{\mathcal{E}}$ it is convenient to write equations
(\ref{induc1}) and (\ref{navie1}) in Fourier space:
 \begin{eqnarray}
\left(\frac{\partial}{\partial t}+\eta z^{2}\right)\hat{b}_{j} & = & \mathrm{i}z_{l}\int\left[\widehat{m}_{j}(\mathbf{z-q)}\hat{\left(\frac{\overline{B}_{l}}{\rho}\right)}(\mathbf{q})-\widehat{m}_{l}(\mathbf{z-q)}\hat{\left(\frac{\overline{B}_{j}}{\rho}\right)}(\mathbf{q})\right]\mathrm{{d}}\mathbf{q}\label{induc2}\\
 & + & \mathrm{i}z_{l}\int\left[\widehat{b}_{l}(\mathbf{z-q)}\hat{\overline{V}}_{j}(\mathbf{q})-\widehat{b}_{j}(\mathbf{z-q)}\hat{\overline{V}}_{l}(\mathbf{q})\right]\mathrm{{d}}\mathbf{q}+\widehat{{\mathfrak{G}}}_{j}.\nonumber \\
\left(\frac{\partial}{\partial t}+\nu z^{2}+\mathrm{i}\nu\left(\mathbf{Gz}\right)\right)\hat{m}_{i} & = & \hat{f}_{i}+\hat{\mathfrak{F}}_{i}-2\left(\mathbf{\Omega}\hat{\mathbf{z}}\right)\left(\hat{\mathbf{z}}\times\hat{\mathbf{m}}\right)_{i}\label{navie2}\\
 & - & \mathrm{i}\pi_{if}(\mathbf{z)}z_{l}\int\left[\widehat{m}_{l}(\mathbf{z-q)}\hat{\overline{V}}_{f}(\mathbf{q})+\widehat{m}_{f}(\mathbf{z-q)}\hat{\overline{V}}_{l}(\mathbf{q})\right]\mathrm{{d}}\mathbf{q}\nonumber \\
 & + & \frac{\mathrm{i}}{\mu}\pi_{if}(\mathbf{z)}z_{l}\int\left[\widehat{b}_{l}(\mathbf{z-q)}\hat{\overline{B}}_{f}\left({\mathbf{q}}\right)+\widehat{b}_{f}(\mathbf{z-q)}\hat{\overline{B}}_{l}\left(q\right)\right]\mathrm{{d}}\mathbf{q},\nonumber \end{eqnarray}
where the turbulent pressure was excluded from (\ref{navie1}) by
convolution with the projection tensor $\pi_{ij}(\mathbf{z)}=\delta_{ij}-\hat{z}_{i}\hat{z}_{j}$,
$\delta_{ij}$ is the Kronecker symbol and $\hat{\mathbf{z}}$ is
a unit wave vector. The equations for the second-order moments which
make contributions to the MEMF can be found from (\ref{induc2},\ref{navie2}).
As the preliminary step we write the equations for the second-order
products of the fluctuating fields, and make the ensemble averaging
of them,
 \begin{eqnarray}
\frac{\partial}{\partial t}\left\langle \hat{m}_{i}\left(\mathbf{z}\right)\hat{b}_{j}\left(\mathbf{z}'\right)\right\rangle  & = & Th_{ij}^{\varkappa}(\mathbf{z},\mathbf{z'})-\left(\eta z'^{2}+\nu z^{2}+\mathrm{i}\nu\left(\mathbf{Gz}\right)\right)\left\langle \hat{m}_{i}\left(\mathbf{z}\right)\hat{b}_{j}\left(\mathbf{z}'\right)\right\rangle \label{eq:kappa1}\\
 &  & \mathrm{i}z'_{l}\int\left[\left\langle \hat{m}_{i}\left(\mathbf{z}\right)\hat{m}_{j}(\mathbf{z'-q)}\right\rangle \hat{\left(\frac{\overline{B}_{l}}{\rho}\right)}(\mathbf{q})-\right.\nonumber \\
 & - & \left.\left\langle \hat{m}_{i}\left(\mathbf{z}\right)\hat{m}_{l}(\mathbf{z'-q)}\right\rangle \hat{\left(\frac{\overline{B}_{j}}{\rho}\right)}(\mathbf{q})\right]\mathd\mathbf{q}-2\left(\mathbf{\Omega}\hat{\mathbf{z}}\right)\varepsilon_{iln}\hat{z_{l}}\left\langle \hat{m_{n}}(\mathbf{z})\hat{b}_{j}(\mathbf{z'})\right\rangle \nonumber \\
 & + & \mathrm{i}z'_{l}\int\left[\left\langle \hat{m}_{i}\left(\mathbf{z}\right)\widehat{b}_{l}(\mathbf{z'-q)}\right\rangle \hat{\overline{V}}_{j}(\mathbf{q})-\left\langle \hat{m}_{i}\left(\mathbf{z}\right)\widehat{b}_{j}(\mathbf{z'-q)}\right\rangle \hat{\overline{V}}_{l}(\mathbf{q})\right]\mathrm{{d}}\mathbf{q}\nonumber \\
 & - & \mathrm{i}\pi_{if}(\mathbf{z)}z_{l}\int\left[\left\langle \widehat{m}_{l}(\mathbf{z-q)}\hat{b}_{j}\left(\mathbf{z}'\right)\right\rangle \hat{\overline{V}}_{f}(\mathbf{q})+\left\langle \widehat{m}_{f}(\mathbf{z-q)}\hat{b}_{j}\left(\mathbf{z}'\right)\right\rangle \hat{\overline{V}}_{l}(\mathbf{q})\right]\mathrm{{d}}\mathbf{q}\nonumber \\
 & + & \frac{\mathrm{i}}{\mu}z_{l}\pi_{if}(\mathbf{z)}\int\left[\left\langle \widehat{b}_{l}(\mathbf{z-q)}\hat{b}_{j}\left(\mathbf{z}'\right)\right\rangle \overline{B}_{f}\left(\mathbf{q}\right)+\left\langle \widehat{b}_{f}(\mathbf{z-q)}\hat{b}_{j}\left(\mathbf{z}'\right)\right\rangle \overline{B}_{l}\left(q\right)\right]\mathd\mathbf{q},\nonumber
 \end{eqnarray}
 \begin{eqnarray}
\frac{\partial}{\partial t}\left\langle \hat{m}_{i}\left(\mathbf{z}\right)\hat{m}_{j}\left(\mathbf{z}'\right)\right\rangle  & = & -2\left(\mathbf{\Omega}\hat{\mathbf{z}}\right)\varepsilon_{iln}\hat{z_{l}}\left\langle \hat{m_{n}}(\mathbf{z})\hat{m}_{j}(\mathbf{z'})\right\rangle -2\left(\mathbf{\Omega}\hat{\mathbf{z}}'\right)\varepsilon_{jln}\hat{z_{l}'}\left\langle \hat{m_{i}}(\mathbf{z})\hat{m}_{n}(\mathbf{z'})\right\rangle \label{secm1}\\
 & - & \mathrm{i}\pi_{if}(\mathbf{z)}z_{l}\int\left[\left\langle \widehat{m}_{l}(\mathbf{z-q)}\hat{m}_{j}\left(\mathbf{z}'\right)\right\rangle \hat{\overline{V}}_{f}(\mathbf{q})+\left\langle \widehat{m}_{f}(\mathbf{z-q)}\hat{m}_{j}\left(\mathbf{z}'\right)\right\rangle \hat{\overline{V}}_{l}(\mathbf{q})\right]\mathrm{{d}}\mathbf{q}\nonumber \\
 & - & \mathrm{i}\pi_{jf}(\mathbf{z')}z'_{l}\int\left[\left\langle \hat{m}_{i}\left(\mathbf{z}\right)\widehat{m}_{l}(\mathbf{z-q)}\right\rangle \hat{\overline{V}}_{f}(\mathbf{q})+\left\langle \hat{m}_{i}\left(\mathbf{z}\right)\widehat{m}_{f}(\mathbf{z-q)}\right\rangle \hat{\overline{V}}_{l}(\mathbf{q})\right]\mathrm{{d}}\mathbf{q}\nonumber \\
 & + & \frac{\mathrm{i}}{\mu}\pi_{if}\left(\mathbf{z}\right)z_{l}\int\left[\left\langle \widehat{b}_{l}(\mathbf{z-q)}\hat{m}_{j}\left(\mathbf{z}'\right)\right\rangle \hat{\overline{B}}_{f}\left({\mathbf{q}}\right)+\left\langle \widehat{b}_{f}(\mathbf{z-q)}\hat{m}_{j}\left(\mathbf{z}'\right)\right\rangle \hat{\overline{B}}_{l}\left(q\right)\right]\mathrm{{d}}\mathbf{q}\nonumber \\
 & + & \frac{\mathrm{i}}{\mu}\pi_{jf}(\mathbf{z')}z'_{l}\int\left[\left\langle \hat{m}_{i}\left(\mathbf{z}\right)\widehat{b}_{l}(\mathbf{z-q)}\right\rangle \hat{\overline{B}}_{f}\left({\mathbf{q}}\right)+\left\langle \hat{m}_{i}\left(\mathbf{z}\right)\widehat{b}_{f}(\mathbf{z-q)}\right\rangle \hat{\overline{B}}_{l}\left(q\right)\right]\mathrm{{d}}\mathbf{q}\nonumber \\
 & + &
 Th_{ij}^{v}(\mathbf{z},\mathbf{z'})-\nu\left(z'^{2}+z^{2}+i\left(\mathbf{Gz}\right)+i\left(\mathbf{Gz'}\right)\right)\left\langle
   \hat{m}_{i}\left(\mathbf{z}\right)\hat{m}_{j}\left(\mathbf{z}'\right)\right\rangle
 ,\nonumber
 \end{eqnarray}
 \begin{eqnarray}
\frac{\partial}{\partial t}\left\langle \hat{b}_{i}\left(\mathbf{z}\right)\hat{b}_{j}\left(\mathbf{z}'\right)\right\rangle  & = & Th_{ij}^{h}(\mathbf{z},\mathbf{z'})-\left(\eta z'^{2}+\eta z^{2}\right)\left\langle \hat{b}_{i}\left(\mathbf{z}\right)\hat{b}_{j}\left(\mathbf{z}'\right)\right\rangle \label{eq:mag1}\\
 & + & \mathrm{i}z'_{l}\int\left[\left\langle \hat{b}_{i}\left(\mathbf{z}\right)\hat{m}_{j}(\mathbf{z'-q)}\right\rangle \hat{\left(\frac{\overline{B}_{l}}{\rho}\right)}(\mathbf{q})-\left\langle \hat{b}_{i}\left(\mathbf{z}\right)\hat{m}_{l}(\mathbf{z'-q)}\right\rangle \hat{\left(\frac{\overline{B}_{j}}{\rho}\right)}(\mathbf{q})\right]\mathd\mathbf{q}\nonumber \\
 & + & \mathrm{i}z_{l}\int\left[\left\langle \hat{m}_{i}(\mathbf{z-q)}\hat{b}_{j}\left(\mathbf{z}'\right)\right\rangle \hat{\left(\frac{\overline{B}_{l}}{\rho}\right)}(\mathbf{q})-\left\langle \hat{m}_{l}(\mathbf{z-q)}\hat{b}_{j}\left(\mathbf{z}'\right)\right\rangle \hat{\left(\frac{\overline{B}_{i}}{\rho}\right)}(\mathbf{q})\right]\mathd\mathbf{q}\nonumber \\
 & + & \mathrm{i}z'_{l}\int\left[\left\langle
     \hat{b}_{i}\left(\mathbf{z}\right)\widehat{b}_{l}(\mathbf{z'-q)}\right\rangle
   \hat{\overline{V}}_{j}(\mathbf{q})-\left\langle
     \hat{b}_{i}\left(\mathbf{z}\right)\widehat{b}_{j}(\mathbf{z'-q)}\right\rangle
   \hat{\overline{V}}_{l}(\mathbf{q})\right]\mathrm{{d}}\mathbf{q}\nonumber\\
 & + & \mathrm{i}z_{l}\int\left[\left\langle
     \hat{b}_{l}\left(\mathbf{z-q}\right)\widehat{b}_{j}(\mathbf{z')}\right\rangle
   \hat{\overline{V}}_{i}(\mathbf{q})-\left\langle
     \hat{b}_{i}\left(\mathbf{z-q}\right)\widehat{b}_{j}(\mathbf{z')}\right\rangle
   \hat{\overline{V}}_{l}(\mathbf{q})\right]\mathrm{{d}}\mathbf{q},
\nonumber \end{eqnarray}
where, the terms $Th_{ij}^{(\varkappa,v,h)}$ involve the third-order
moments of fluctuating fields and second-order moments of them with
the forcing term. 

To proceed further, it is convenient to introduce some notations
which are used in the literature. The double Fourier transformation
of an ensemble average of two fluctuating quantities, say $f$ and
$g$, taken at equal times and at the different positions $\mathbf{x},\,\mathbf{x}'$,
is given by
 \begin{equation}
\left\langle f\left(\mathbf{x}\right)g\left(\mathbf{x}'\right)\right\rangle =\int\int\left\langle \hat{f}\left(\mathbf{z}\right)\hat{g}\left(\mathbf{z}'\right)\right\rangle e^{\mathrm{i}\left(\mathbf{z}\cdot\mathbf{x}+\mathbf{z}'\cdot\mathbf{x}'\right)}\mathd^{3}\mathbf{z}\mathd^{3}\mathbf{z}'.\label{cor1}\end{equation}
Let us define the ``fast'' spatial variable $\mathbf{r}$  by the
relative difference of $\mathbf{x},\mathbf{x}'$ coordinates, 
$\mathbf{r}=\mathbf{x}-\mathbf{x}'$. The ``slow'' spatial variable
$\mathbf{R}$  is  $\mathbf{R}=\left(\mathbf{x}+\mathbf{x}'\right)/2$. Then, eq. (\ref{cor1}) can be written in the form
 \begin{eqnarray}
\left\langle f\left(\mathbf{x}\right)g\left(\mathbf{x}'\right)\right\rangle  & = & \int\int\left\langle \hat{f}\left(\mathbf{k}+\frac{1}{2}\mathbf{K}\right)\hat{g}\left(-\mathbf{k}+\frac{1}{2}\mathbf{K}\right)\right\rangle \mathe^{\mathrm{i}\left(\mathbf{K}\cdot\mathbf{R}+\mathbf{k}\cdot\mathbf{r}\right)}\mathd^{3}\mathbf{K}\mathd^{3}\mathbf{k},\label{cor1b}\end{eqnarray}
where I have introduced two wave vectors: $\mathbf{k}=\left(\mathbf{z}-\mathbf{z}'\right)/2$
and $\mathbf{K}=\mathbf{z}+\mathbf{z}'$. Following BS05, we
define the correlation function of $\hat{\mathbf{f}}$ and $\hat{\mathbf{g}}$
obtained from (\ref{cor1b}) by integration with respect to $\mathbf{K}$,
\begin{equation}
\Phi\left(\hat{f},\hat{g},\mathbf{k},\mathbf{R}\right)=\int\left\langle \hat{f}\left(\mathbf{k}+\frac{1}{2}\mathbf{K}\right)\hat{g}\left(-\mathbf{k}+\frac{1}{2}\mathbf{K}\right)\right\rangle \mathe^{\mathrm{i}\left(\mathbf{K}\cdot\mathbf{R}\right)}\mathd^{3}\mathbf{K}.\label{eq:cor1a}\end{equation}
For further convenience we define the second order correlations of
momentum density, magnetic fluctuations and the cross-correlations
of momentum and magnetic fluctuations via
 \begin{eqnarray}
\hat{v}_{ij}\left(\mathbf{k},\mathbf{R}\right) & = & \Phi(\hat{m}_{i},\hat{m}_{j},\mathbf{k},\mathbf{R}),\bar{\rho}^{2}\left\langle u^{2}\right\rangle \left(\mathbf{R}\right)=\int\hat{v}_{ii}\left(\mathbf{k},\mathbf{R}\right)\mathd^{3}\mathbf{k},\\
\hat{h}_{ij}\left(\mathbf{k},\mathbf{R}\right) & = & \Phi(\hat{b}_{i},\hat{b}_{j},\mathbf{k},\mathbf{R}),\left\langle b^{2}\right\rangle \left(\mathbf{R}\right)=\int\hat{h}_{ii}\left(\mathbf{k},\mathbf{R}\right)\mathd^{3}\mathbf{k},\label{cor2}\\
\hat{\varkappa}_{ij}\left(\mathbf{k},\mathbf{R}\right) & = & \Phi(\hat{m}_{i},\hat{b}_{j},\mathbf{k},\mathbf{R}),\bar{\rho}\mathcal{E}_{i}\left(\mathbf{R}\right)=\varepsilon_{ijk}\int\hat{\varkappa}_{jk}\left(\mathbf{k},\mathbf{R}\right)\mathd^{3}\mathbf{k}.\end{eqnarray}

Let us now return to equations (\ref{eq:kappa1}), (\ref{secm1}) and
(\ref{eq:mag1}). As the first step, we approximate the
$Th_{ij}^{(\varkappa,v,h)}$ terms by the corresponding $\tau$ relaxation terms of the second-order
contributions,
 \begin{align}
Th_{ij}^{(\varkappa)} & \rightarrow-\left\langle \hat{m}_{i}\left(\mathbf{z}\right)\hat{b}_{j}\left(\mathbf{z}'\right)\right\rangle /\tau_{c},\label{eq:thx}\\
Th_{ij}^{(v)} & \rightarrow-\frac{\left\langle \hat{m}_{i}\left(\mathbf{z}\right)\hat{m}_{j}\left(\mathbf{z}'\right)\right\rangle -\left\langle \hat{m}_{i}\left(\mathbf{z}\right)\hat{m}_{j}\left(\mathbf{z}'\right)\right\rangle ^{\left(0\right)}}{\tau_{c}},\label{eq:thv}\\
Th_{ij}^{(h)} & \rightarrow-\frac{\left\langle \hat{b}_{i}\left(\mathbf{z}\right)\hat{b}_{j}\left(\mathbf{z}'\right)\right\rangle -\left\langle \hat{b}_{i}\left(\mathbf{z}\right)\hat{b}_{j}\left(\mathbf{z}'\right)\right\rangle ^{(0)}}{\tau_{c}},\label{eq:thh}\end{align}
where the superscript $^{\left(0\right)}$ denotes the moments of
the background turbulence. Here, $\tau_{c}$ is independent on $\mathbf{k}$
and it is independent on the mean fields as well.
Furthermore, for the sake of simplicity, we restrict
ourselves to the high Reynolds numbers limit and discard the microscopic
diffusion terms. As the next step we make the Taylor expansion with
respect to the ``slow'' variables and take the Fourier transformation,
(\ref{eq:cor1a}), about them. The details of this procedure can be
found in BS05. In result, we obtain equations for the second order
correlations of momentum density, magnetic fluctuations and the cross-correlations
of momentum and magnetic fluctuations,

\begin{eqnarray}
\frac{\partial\hat{\varkappa}_{ij}}{\partial t} & = & 
-\mathrm{i}\left(\mathbf{\overline{B}k}\right)\left(\frac{\hat{v}_{ij}}{\rho}
-\frac{\hat{h}_{ij}}{\mu}\right)
+\frac{\left(\overline{\mathbf{B}}\nabla\right)}{2}\left(\frac{\hat{v}_{ij}}{\rho}
+\frac{\hat{h}_{ij}}{\mu}\right)+\frac{\left(\overline{\mathbf{B}}\mathbf{k}\right)}{2\rho}G_{s}\frac{\partial\hat{v}_{ij}}{\partial
k^{s}}
-\frac{\left(\mathbf{G\overline{B}}\right)}{2\rho}\hat{v}_{ij}\label{eq:secm2a}\\
 & + &
 \frac{1}{\rho}G_{l}\hat{v}_{il}B_{j}+\frac{\hat{h}_{lj}\overline{B}_{i,l}}{\mu}
-\frac{\hat{v}_{il}\overline{B}_{j,l}}{\rho}
-\frac{k_{l}\overline{B}_{l,f}}{2}\frac{\partial}{\partial
  k_{f}}\left[\frac{\hat{v}_{ij}}{\rho}
+\frac{\hat{h}_{ij}}{\mu}\right]-\frac{2}{\mu}\hat{k}_{i}\hat{k}_{f}\overline{B}_{f,l}\hat{h}_{lj}\nonumber \\
 & + &
 \overline{V}_{j,l}\hat{\varkappa}_{il}-\overline{V}_{i,l}\hat{\varkappa}_{lj}
+2\hat{k}_{i}\hat{k}_{f}\hat{\varkappa}_{lj}\overline{V}_{f,l}
+k_{l}\overline{V}_{f,l}\frac{\partial\hat{\varkappa}_{ij}}{\partial k_{f}}
- \frac{\hat{\varkappa}_{ij}}{\tau_{c}}
-2\left(\Omega\hat{k}\right)\hat{k}_{p}\varepsilon_{ipl}\hat{\varkappa}_{lj}\nonumber \\
&-& 2\frac{\mathrm{i}}{k}\left(\Omega\hat{k}\right)\hat{k}_{p}\varepsilon_{ipl}\left(\hat{k}\nabla\right)\hat{\varkappa}_{lj}
 + \frac{\mathrm{i}}{k}\varepsilon_{ipl}\left(\left(\Omega\hat{k}\right)\nabla_{p}\hat{\varkappa}_{lj}+\hat{k}_{p}\left(\Omega\nabla\right)\hat{\varkappa}_{lj}\right)
,\nonumber \end{eqnarray}
\begin{eqnarray}
\frac{\partial\hat{v}_{ij}}{\partial t} & = & -2\left(\Omega\hat{k}\right)\hat{k}_{p}\left(\varepsilon_{ipl}\hat{v}_{lj}+\varepsilon_{jpl}\hat{v}_{il}\right)-\frac{\hat{v}_{ij}-\hat{v}_{ij}^{(0)}}{\tau_{c}}-\hat{v}_{lj}\overline{V}_{i,l}-\hat{v}_{il}\overline{V}_{j,l}\label{secm2}\\
 & + &
 2\hat{k}_{f}\overline{V}_{f,l}\left(\hat{k}_{i}\hat{v}_{lj}+\hat{k}_{j}\hat{v}_{il}\right)
+k_{l}\overline{V}_{f,l}\frac{\partial \hat{v}_{ij}}{\partial
  k_{f}}
-\mathrm{i}\left({\mathbf{\overline{B}k}}\right)\left(\hat{\varkappa}_{ij}-\hat{\varkappa}_{ji}^{\ast}\right)\nonumber \\
 & + & \frac{1}{2}\overline{B}_{l}\left(\hat{\varkappa}_{ij,l}+\hat{\varkappa}_{ji,l}^{\ast}\right)+\overline{B}_{i,l}\hat{\varkappa}_{jl}^{\ast}+\overline{B}_{j,l}\hat{\varkappa}_{i,l}-2\hat{k}_{f}\overline{B}_{f,l}\left(\hat{k}_{i}\hat{\varkappa}_{jl}^{\ast}+\hat{k}_{j}\hat{\varkappa}_{i,l}\right)\nonumber \\
 & - & \frac{\overline{B}_{l,f}}{2}k_{l}\frac{\partial}{\partial k_{f}}\left(\hat{\varkappa}_{ij}+\hat{\varkappa}_{ji}^{\ast}\right)+\frac{\mathrm{i}}{k}\varepsilon_{ipl}\left[\hat{k}_{p}\left(\left(\Omega\nabla\right)-2\left(\Omega\hat{k}\right)\left(\hat{\mathbf{k}}\nabla\right)\right)\right.\nonumber \\
 & + &
 \left.\left(\Omega\hat{k}\right)\nabla_{p}\right]\left(\varepsilon_{ipl}\hat{v}_{lj}
-\varepsilon_{jpl}\hat{v}_{il}\right),\nonumber \end{eqnarray}
\begin{eqnarray}
\frac{\partial\hat{h}_{ij}}{\partial t} & = & -\frac{\hat{h}_{ij}-\hat{h}_{ij}^{(0)}}{\tau_{c}}+\hat{h}_{il}\overline{V}_{j,l}+\hat{h}_{lj}\overline{V}_{i,l}+k_{l}\overline{V}_{f,l}\frac{\partial\hat{h}_{ij}}{\partial k_{f}}+\frac{\mathrm{i}\left({\mathbf{\overline{B}k}}\right)}{\rho}\left(\hat{\varkappa}_{ij}-\hat{\varkappa}_{ji}^{\ast}\right)\label{eq:secm2b}\\
 & + & \left\{ \frac{\left(\overline{B}\nabla\right)}{2\rho}-\frac{\left({\mathbf{\overline{B}G}}\right)}{2\rho}\right\} \left(\hat{\varkappa}_{ij}+\hat{\varkappa}_{ji}^{\ast}\right)-\left(\frac{\overline{B}_{j}}{\rho}\right)_{,l}\hat{\varkappa}_{li}^{\ast}-\left(\frac{\overline{B}_{i}}{\rho}\right)_{,l}\hat{\varkappa}_{lj}\nonumber \\
 & - & \frac{1}{2}\left(\frac{\overline{B}_{l}}{\rho}\right)_{,f}k_{l}\frac{\partial\left(\hat{\varkappa}_{ij}+\hat{\varkappa}_{ji}^{\ast}\right)}{\partial k_{f}},\nonumber \end{eqnarray}
where
 $\hat{\varkappa}_{ji}^{\ast}=\Phi(\hat{b}_{j},\hat{m}_{i},\mathbf{k},\mathbf{R})$,
$\mathbf{\hat{k}}$ is the unit wave vector, the indexes behind the
comma stand for the spatial derivatives. Equations 
(\ref{eq:secm2a},\ref{secm2},\ref{eq:secm2b}) are in agreement
with those considered in the paper by \citet{kle-rog:04c}. 

To solve (\ref{eq:secm2a},\ref{secm2},\ref{eq:secm2b}) we neglect
the time derivatives at the left hand side of equations and apply
the perturbation method. The mean field inhomogeneities and stratification
scales of turbulence are considered as small. We shall not reproduce
explicitly the rather bulky derivations which are explained elsewhere:
Rogachevskii \& Kleeorin\citeyearpar{kle-rog:04a,kle-rog:04b}. The
solution of (\ref{eq:secm2a},\ref{secm2},\ref{eq:secm2b}) will
be given for two specific cases. In the first case we apply no restriction
to the angular velocity ( the Coriolis number, $\Omega^{\ast}=2\Omega\tau_{c}$,
is arbitrary) and LSMF is assumed to be weak. In the second case we
keep the linear terms in angular velocity and solve eqs.(\ref{eq:secm2a},\ref{secm2},\ref{eq:secm2b})
for the case of arbitrary $\beta=\bar{B}/\left(u_{c}\sqrt{\mu\rho}\right)$,
where $\bar{B}$ is the strength of the LSMF. In all derivations we
keep contributions which are the first order in the shear. Furthermore,
for the contributions involving the shear we make two additional simplifications.
The first one is that we neglect the density stratification, but leave
the contributions of the turbulence intensity stratification. Additionally,
we discard the joint effect of the Coriolis force and the shear to
the MEMF.  In the present study I consider  an intermediate
nonlinearity which implies that effect of the mean magnetic field and
global rotation is not enough strong in order to affect the
correlation time of turbulent velocity field.

For integration in $\mathbf{k}$-space I adopt the quasi-isotropic form of the spectra
\citep{rob-saw,kit-rud:1993b} for the background turbulence.
Additionally, the background magnetic fluctuations are helical, while
there is no prescribed kinetic helicity in the background turbulence:
\begin{eqnarray}
\hat{v}_{ij}^{(0)} & = & \left\{ \pi_{ij}\left(\mathbf{k}\right)+\frac{\mathrm{i}}{2k^{2}}\left(k_{i}\nabla_{j}-k_{j}\nabla_{i}\right)\right\} \frac{\rho^{2}E\left(k,\mathbf{R}\right)}{8\pi k^{2}},\label{eq:spectr1}\\
\hat{h}_{ij}^{(0)} & = & \left\{ \left(\pi_{ij}\left(\mathbf{k}\right)+\frac{\mathrm{i}}{2k^{2}}\left(k_{i}\nabla_{j}-k_{j}\nabla_{i}\right)\right)\frac{\mathcal{B}\left(k,\mathbf{R}\right)}{8\pi k^{2}}-\mathrm{i}\varepsilon_{ijp}k_{p}\frac{\mathcal{N}\left(k,\mathbf{R}\right)}{8\pi k^{4}}\right\} ,\label{eq:spectr2}\end{eqnarray}
where, the spectral functions $E(k,\mathbf{R}),\mathcal{B}(k,\mathbf{R}),\mathcal{N}(k,R)$
define, respectively, the intensity of the velocity fluctuations,
the intensity of the magnetic fluctuations and amount of current helicity
in the background turbulence. They are defined via 
\begin{eqnarray}
\left\langle u^{(0)2}\right\rangle  & = & \int\frac{E\left(k,\mathbf{R}\right)}{4\pi k^{2}}\mathd^{3}\mathbf{k},\,\,\left\langle b^{(0)2}\right\rangle =\int\frac{\mathcal{B}\left(k,\mathbf{R}\right)}{4\pi k^{2}}\mathd^{3}\mathbf{k},\,\, h_{C}^{\left(0\right)}=\frac{1}{\mu\rho}\int\frac{\mathcal{N}\left(k,\mathbf{R}\right)}{4\pi k^{2}}\mathd^{3}\mathbf{k},\label{eq:spectr3}\end{eqnarray}
where $h_{\mathcal{C}}^{\left(0\right)}={\displaystyle \left\langle \mathbf{b^{\left(0\right)}\cdot\nabla\times b^{\left(0\right)}}\right\rangle /\left(\mu\rho\right)}$.
In final results we use the relation between intensities of magnetic
and kinetic fluctuations which is defined via $\mathcal{B}\left(k,\mathbf{R}\right)=\varepsilon\mu\bar{\rho}E\left(k,\mathbf{R}\right)$.
The state with $\varepsilon=1$ means equipartition between energies
of magnetic an kinetic fluctuations in the background turbulence. 
The point to note is that inconsistency between (\ref{eq:spectr1})
and (\ref{eq:spectr2}) does not influence the final results. 
The general structure of the mean electromotive force vector
obtained within the given framework are in agreement with the known
results from the literature \citep{rad-kle-rog,kle-rog:04a}. We keep
the current helicity contribution in the background turbulence to
investigate the nonlinear saturation phase of the helical large-scale
dynamo. 

The final remarks in this section concern with discussion given in the
paper by \citet{rad-rhe}. There, authors argue that $\tau$ approximation
may lead to  results which are in conflict with those of SOCA.
One difference is apparent between the two approaches: there is no overlap
in applicability limits of SOCA and $\tau$ approximation.
The given scheme to obtain
(\ref{eq:secm2a},\ref{secm2},\ref{eq:secm2b}) is hardly justified 
for small hydrodynamic Reynolds numbers. The same is true in a highly
conductivity limit, where SOCA can be valid only for the small
Strouhal numbers. Currently, the range of $\tau$ approximation validity
is purely understood. This problem requires further careful study.

There is another reason for difference between results presented  in
the paper and those of SOCA. In the given variant of $\tau$ approximation the
relaxation time $\tau_c$ is independent of $\mathbf{k}$. This issue is
especially important in computing effects of the nonuniform LSMF and
shear. Perhaps, the spectral $\tau$-approximation can correct this
defect. For more detail, see  \citep{rad-kle-rog,kle-rog:07p}.  
Hense, in confronting MTA and SOCA, it is of some use to simplify the expressions
obtained within SOCA by applying the mixing-length approximation.
The transition from SOCA to MLT can be done by replacing the spectrum
of turbulent fields by the single-scaled function of the form 
$\delta\left(k-\ell_{c}^{-1}\right)\delta\left(\omega\right)$,
and applying $\eta k^{2}=\nu k^{2}=\tau_{c}^{-1}$, here $\ell_{c}$
is the correlation length of the turbulence. 
For more details, see \citep{kit:1991,kit-pip-rud}.

\section{Results}

\subsection{Weak LSMF, arbitrary Coriolis number}

\subsubsection{Spatially uniform LSMF}

We divide the electromotive
force into  different contributions, in particular, $\mathbf{\mathcal{E}}^{(a)}$
contains the effects of stratification, and $\mathbf{\mathcal{E}}^{(s)}$
is due to shear. The contributions due to shear are computed only
in slow rotation limit. We find the following expression for
 $\mathbf{\mathcal{E}}^{(a)}$:
\begin{eqnarray}
\mathbf{\mathcal{E}}^{(a)} & = & \left\{ \left(\varepsilon-1\right)\left(f_{2}^{(a)}\left(\mathbf{U\times}\overline{\mathbf{B}}\right)+f_{1}^{(a)}\left(\mathbf{e}\cdot\mathbf{\overline{B}}\right)\left(\mathbf{e\times}\mathbf{U}\right)\right)+f_{3}^{(a)}\left(\mathbf{G\times\overline{B}}\right)\right.\label{emfa1}\\
 & + & f_{1}^{(a)}\left(\left(\mathbf{e}\cdot\mathbf{G}\right)\left(\mathbf{e\times\overline{B}}\right)+\left(\varepsilon-2\right)\left(\mathbf{e}\cdot\mathbf{\overline{B}}\right)\left(\mathbf{e\times G}\right)\right)\nonumber \\
 & + & f_{4}^{(a)}\mathbf{e}\left(\mathbf{e}\cdot\mathbf{\overline{B}}\right)\left(\mathbf{e}\cdot\mathbf{U}\right)+f_{11}^{(a)}\overline{\mathbf{B}}\left(\mathbf{e}\cdot\mathbf{U}\right)+f_{5}^{(a)}\mathbf{e}\left(\mathbf{e}\cdot\overline{\mathbf{B}}\right)\left(\mathbf{e}\cdot\mathbf{G}\right)\nonumber \\
 & + & f_{8}^{(a)}\left(\mathbf{e}\left(\mathbf{\overline{B}}\cdot\mathbf{U}\right)+\mathbf{U}\left(\mathbf{e}\cdot\mathbf{\overline{B}}\right)\right)+f_{6}^{(a)}\left(\mathbf{e}\left(\mathbf{\overline{B}}\cdot\mathbf{G}\right)+\mathbf{G}\left(\mathbf{e}\cdot\mathbf{\overline{B}}\right)\right)+f_{10}^{(a)}\overline{\mathbf{B}}\left(\mathbf{e}\cdot\mathbf{G}\right)\nonumber \\
 & + & \left.f_{9}^{(a)}\left(\mathbf{e}\left(\mathbf{\overline{B}}\cdot\mathbf{U}\right)-\mathbf{U}\left(\mathbf{e}\cdot\overline{\mathbf{B}}\right)\right)+f_{7}^{(a)}\left(\mathbf{e}\left(\mathbf{\overline{B}}\cdot\mathbf{G}\right)-\mathbf{G}\left(\mathbf{e}\cdot\mathbf{\overline{B}}\right)\right)\right\} \left\langle u^{(0)2}\right\rangle \tau_{c}\nonumber \\
 & + & 2\left\{ f_{2}^{(a)}\mathbf{\overline{B}}-f_{1}^{(a)}\mathbf{e}\left(\mathbf{e}\cdot\overline{\mathbf{B}}\right)\right\} \tau_{c}h_{\mathcal{C}}^{\left(0\right)},\nonumber \end{eqnarray}
 where functions $f_{\{ n\}}^{(a)}=f_{\{ n\}}^{(a)}\left(\Omega^{\ast},\varepsilon\right)$
(and all which are used below) are given in Appendix A, $\mathbf{U}=\mathbf{\nabla}\log\left\langle u^{(0)2}\right\rangle $
is a scale of the turbulence intensity stratification, $\mathbf{e}=\mathbf{\mathbf{\Omega}}/|\Omega|$
is a unit vector in direction of global rotation. For the slow rotation
limit ($\Omega^{\ast}\rightarrow0$) we get :
\begin{eqnarray}
\mathcal{E}^{(a)}|_{\Omega^{\ast}\rightarrow0} & = & \alpha\circ\overline{\mathbf{B}}+\left\langle u^{(0)2}\right\rangle \tau_{c}\left\{ \frac{(\varepsilon-1)}{6}\left(\mathbf{U\times\overline{B}}\right)+\frac{\varepsilon}{6}\left(\mathbf{G\times\overline{B}}\right)\right\} \label{slow1}\\
 & + & \left\langle u^{(0)2}\right\rangle \tau_{c}\frac{\Omega^{\ast}}{12}\left\{ \left(\varepsilon+2\right)\left(\left(\mathbf{\left(G\times e\right)\times\overline{B}}\right)\right)+\left(\varepsilon+1\right)\left(\left(\mathbf{\left(U\times e\right)\times\overline{B}}\right)\right)\right\} ,\nonumber \\
\alpha_{ij} & =\!\! & \delta_{ij}\tau_{c}\left(\left\langle u^{(0)2}\right\rangle \left\{ \frac{2\varepsilon\left(\left(\mathbf{e}\cdot\mathbf{U}\right)+\left(\mathbf{e}\cdot\mathbf{G}\right)\right)\Omega^{\ast}}{15}\!-\!\frac{2\left(\mathbf{e}\cdot\mathbf{U}\right)\Omega^{\ast}}{5}\!-\!\frac{4\left(\mathbf{e}\cdot\mathbf{G}\right)\Omega^{\ast}}{5}\right\} \!\!+\!\!\frac{h_{\mathcal{C}}^{(0)}}{3}\right)\nonumber \\
 & + & \tau_{c}\left\langle u^{(0)2}\right\rangle \frac{\Omega^{\ast}}{20}\left\{ \left(e_{i}G_{j}+e_{j}G_{i}\right)\left(\varepsilon+4\right)+\left(e_{i}U_{j}+e_{j}U_{i}\right)\left(\varepsilon+\frac{11}{3}\right)\right\} ,\label{alpsl}\end{eqnarray}
where only linear terms in $\Omega$ are kept. Except contributions
due to $\mathbf{G}$ equations (\ref{slow1}) and (\ref{alpsl}) are
in agreement with results by \citet{rad-kle-rog} and \citet{bra-sub:04}.
The mean transport of the LSMF due to stratification of turbulence
is given by second term in (\ref{slow1}). They are in agreement with
the mixing-length expressions obtained by \citet{kit:1991}. Note
that, additional components of the turbulent transport may be excited
due to the antisymmetric part of $\alpha$-tensor in (\ref{alpsl}).

For the fast rotation limit ($\Omega^{\ast}\rightarrow\infty$) of
(\ref{emfa1}) we get
 \begin{eqnarray}
\mathbf{\mathcal{E}}^{(a)}|_{\Omega^{\ast}\rightarrow\infty} & \rightarrow & \frac{\pi\tau_{c}}{2}\left(\frac{h_{\mathcal{C}}^{\left(0\right)}}{2\Omega^{\ast}}-\left\langle u^{(0)2}\right\rangle \left(\frac{\left(\mathbf{e}\cdot\mathbf{U}\right)}{2}+\left(\mathbf{e}\cdot\mathbf{G}\right)\right)\right)\left(\overline{\mathbf{B}}-\mathbf{e}\left(\mathbf{e}\cdot\overline{\mathbf{B}}\right)\right),\label{frot}\end{eqnarray}
where, we keep the next order contribution in $\Omega^{\ast}$ for
the current helicity, as well. The reason for this will be clarified
later in section \ref{sec:The-current-helicity}. Except the helicity
term, eq.(\ref{frot}) is in identical agreement \emph{}with the mixing-length
approximation results obtained by \citet{kit-rud:1993b} within SOCA.

\begin{figure}
\begin{centering}\includegraphics[width=0.5\linewidth]{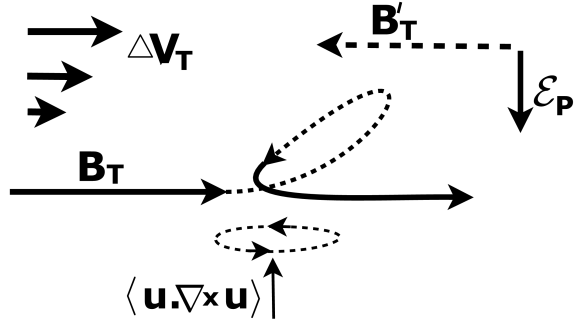}\par\end{centering}

\caption{\label{cap:1}The modification of standard alpha effect (cf. \citet{krarad80})
due to shear. The helical motions (denoted with $\left\langle \mathbf{u}\cdot\nabla\times\mathbf{u}\right\rangle $)
go up, drag and twist the LSMF $\mathbf{B}_{T}$, where index $_{T}$
denotes the toroidal component of LSMF. The shear,$\Delta V_{T}$,
additionally, folds the loop in direction of large-scale flow. The
effect is equivalent to inducing the transversal large-scale electromotive
force, $\mathbf{\mathcal{E}}_{P}$ (here index $_{P}$ denotes the
poloidal component of the MEMF), and the magnetic field, $\mathbf{B}'_{T}$
parallel to original one. Direction of the induced field depends on
the sign of the helicity. For the situation given on the picture,
the induced field $\mathbf{B}'_{T}$ quenches the original LSMF in
direction of the gradient of the mean flow. This means that the LSMF
is effectively pumped in opposite direction.}
\end{figure}
In the case of the spatially uniform LSMF the shear contributions
to the mean electromotive force are expressed as follows: 

\begin{eqnarray}
\mathcal{E}_{i}^{(s)}\!\! & =\!\! & \varepsilon_{inm}\left\{ A_{4}U_{k}\overline{B}_{n}\overline{V}_{m,k}\!+\! A_{2}\overline{B}_{k}\overline{V}_{n,k}U_{m}\!+\! A_{3}\left(\mathbf{\overline{B}\cdot U}\right)\overline{V}_{m,n}\!+\! A_{1}\overline{V}_{k,n}\overline{B}_{k}U_{m}\right\} \left\langle u^{(0)2}\right\rangle \label{emfas1}\\
 & + &
 \tau_{c}^{2}\frac{h_{\mathcal{C}}^{\left(0\right)}}{2}\left(\mathbf{W}\times\mathbf{\overline{B}}\right)_i
-\frac{13}{30}\tau_{c}^{2}h_{\mathcal{C}}^{\left(0\right)}
\left\{\overline{V}_{n,i}+\overline{V}_{i,n}\right\}\overline{B}_{n} ,\nonumber \end{eqnarray}
where $\mathbf{W}=\nabla\times\mathbf{\overline{V}}$, we assume that
$(\mathbf{U}\cdot\mathbf{\nabla})\mathbf{\bar{V}}=0$ and
 $A_{1}=\left(2\varepsilon-1\right)\tau_{c}^{2}/15$,
$A_{2}=-\left(3\varepsilon+1\right)\tau_{c}^{2}/15$,
$A_{3}={\displaystyle \left(\varepsilon+1\right)\tau_{c}^{2}}/6$,
 $A_{4}=-A_{3}$,. Coefficients
$A_{1-3}$ correspond to those from \citet{rud-ki:06a} (hereafter RK06) and
$A_4$ is corresponding to their $A_5$.  Recently, similar
contributions of the large-scale shear were calculated within SOCA
by \citet{rad-step} (RS06), as well.
We have to note that both the RK06 and RS06 results are related  with the
case $\varepsilon=0$. The (\ref{emfas1}) differs with
results obtained in RK06 and RS06 papers. For example, after applying
the mixing-length relations 
 $\eta k^{2}=\nu k^{2}=\tau_{c}^{-1}$  to expressions given by RK06 we get
$A_{1}=\tau_{c}^{2}/3$ (in our case $-\tau_{c}^{2}/15$) and $A_{2}=-\tau_{c}^{2}/60$
(compare to our $-\tau_{c}^{2}/15$). Unfortunately RK06 did not present
the results for other coefficients. The comparison with RS06 is given
in Appendix B. The difference between the given results and those
by RK06 and RS06  can be explained, in part, by the crudeness of the given version
of tau approximation. Here, we assume that $\tau_{c}$ is independent
of $\mathbf{k}$. This especially influences the accuracy of calculations
of the contributions due to shear because they involve the derivatives
in $\mathbf{k}$ space. 

According to (\ref{emfas1}) the joint effect of current helicity
and shear contributes to pumping of LSMF. The interpretation of the
effect is difficult to illustrate. To show the
general idea we invoke an auxiliary illustration of effect for the
helical turbulent motions. It is shown on Fig.\ref{cap:1}.

\subsubsection{Anisotropic diffusion, the $\Omega\times\mathbf{J}$ and shear-current
effects}

\begin{figure}
\begin{centering}\includegraphics[width=0.5\linewidth]{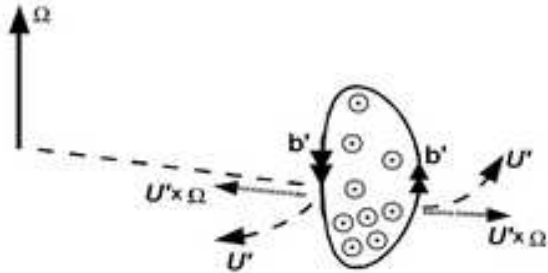}\par\end{centering}

\caption{\label{cap:omxj}An illustration of $\mathbf{\Omega}\times\mathbf{J}$
effect in disk geometry. Direction of rotation is marked by $\mathbf{\Omega}$,
the large-scale toroidal field has opposite direction to rotational
velocity and it is marked by $\odot$, what means that LSMF is perpendicular
to the figure's plane and it is directed to the reader. The loop of
fluctuating magnetic field, $\mathbf{b}'$, comprises LSMF that is
nonuniform along the axis of rotation. Its direction is marked by
double arrows. The small-scale Lorentz forces induce the azimuthal
fluctuations of velocity, $\mathbf{u}'\sim\left(\mathbf{b'\cdot}\nabla\right)\mathbf{\overline{B}}$.
They are marked by dashed lines ending with arrows. The Coriolis force
deflects these fluctuations to radial direction (this is marked by
dotted lines). The resulting electromotive force has the same direction
as the original LSMF and it is proportional to 
$\left\langle b'^{2}\right\rangle \left(\Omega\cdot\nabla\right)\mathbf{\overline{B}}$.}
\end{figure}

In rotating turbulence the magnetic diffusivity become anisotropic
\citep{kit-pip-rud}. The corresponding part of the MEMF reads,
 \begin{eqnarray}
\mathcal{E}_{i}^{(d)} & = & \left\{ \mathrm{f}_{1}^{(d)}e_{n}\overline{B}_{\mathrm{n,i}}+\mathrm{f}_{2}^{(d)}\varepsilon_{\mathrm{inm}}\overline{B}_{\mathrm{m,n}}+\varepsilon\mathrm{f}_{3}^{(d)}e_{i}e_{n}e_{m}\overline{B}_{\mathrm{m,n}}\right.\nonumber \\
 & + & \left.\mathrm{f}_{1}^{(a)}\varepsilon_{inm}e_{n}e_{l}
\left(2\varepsilon\overline{B}_{l,m}-
\left(\varepsilon+1\right)\overline{B}_{m,l}\right)+
\varepsilon\mathrm{f}_{4}^{(d)}e_{n}\overline{B}_{\mathrm{i,n}}\right\} 
\left\langle u^{(0)2}\right\rangle \tau_{c},\label{dif1}\end{eqnarray}
where functions $f_{\{ n\}}^{(d)}=f_{\{ n\}}^{(d)}\left(\Omega^{\ast}\right)$
are given in Appendix A. If we put the magnetic fluctuations in background
turbulence equal to zero in (\ref{dif1}) ($\varepsilon=0$), we return
to results obtained by \citet{kit-pip-rud}. The magnetic fluctuation
contributions in (\ref{dif1}) give rise to the $\mathbf{\Omega}\times\mathbf{J}$
effect (see \citet{rad69,krarad80,rad-kle-rog,kit:03}) and to additions
in anisotropic diffusion. In the slow-rotation limit eq. (\ref{dif1})
can be reduced to
 \begin{eqnarray}
\mathcal{E}_{i}^{(d)}|_{\Omega*\rightarrow0} & = & 
\left\{
  e_{n}\left(\left(\varepsilon+5\right)\overline{B}_{\mathrm{n,i}}
+6\varepsilon\overline{B}_{\mathrm{i,n}}\right)\frac{\Omega^{\ast}}{10}
-\varepsilon_{\mathrm{inm}}\overline{B}_{\mathrm{m,n}}\right\}
 \frac{\left\langle u^{(0)2}\right\rangle \tau_{c}}{3},
\label{dif0}\end{eqnarray}
Eq. (\ref{dif0}) corresponds to results by \citet{bra-sub:04}. Note,
only magnetic fluctuations contribute to the induction term
$\left(\mathbf{e\cdot}\mathbf{\nabla}\right)\mathbf{\overline{B}}$. 
The physical interpretation of this effect is straightforward and
it is shown on Fig.\ref{cap:omxj}

Lets consider the situation in disk geometry and the rotating media
penetrated by the inhomogeneous toroidal LSMF. For simplicity, we
assume that LSMF is nonuniform along the axis of rotation. Let the
direction of LSMF be opposite to direction of rotating plasma. If
the loop of the small-scale fluctuating magnetic field comprises LSMF,
it induces fluctuation of velocity in azimuthal direction. The influence
of the Coriolis force declines the velocities in radial direction.
The effective electromotive force is co-lined with original LSMF and
is proportional to $\left\langle b^{2}\right\rangle \left(\Omega\cdot\nabla\right)\mathbf{\overline{B}}$,
see Fig.\ref{cap:omxj}.

The shear-current effect \citet{kle-rog:04a} (hereafter RK03) is of similar nature
because the large-scale vorticity $\mathbf{W}=\nabla\times\mathbf{\overline{V}}$ and the Coriolis force
act on the turbulent motions in a like manner. The additional contributions
due to shear in the diffusion part of the mean electromotive force
are expressed as follows,

\begin{equation}
\mathcal{E}_{i}^{(V)}=\varepsilon_{inm}
\left\{ C_{2}\overline{B}_{\mathrm{n,l}}\overline{V}_{m\mathrm{,l}}
+C_{1}\overline{V}_{\mathrm{l,m}}\overline{B}_{\mathrm{n,l}}
+C_{3}\overline{V}_{l\mathrm{,m}}\overline{B}_{\mathrm{l,n}}
+C_{4}\overline{B}_{\mathrm{l,n}}\overline{V}_{\mathrm{m,l}}\right\} 
\left\langle u^{(0)2}\right\rangle ,\label{difs1}\end{equation}
where $C_{1}=\left(\varepsilon-3/5\right)\tau_{c}^{2}/6$,
$C_{2}=\left(\varepsilon-1\right)\tau_{c}^{2}/5$,$
C_{3}=\left(1+\varepsilon\right)\tau_{c}^{2}/15$,$C_{4}=-\left(7\varepsilon+11\right)\tau_{c}^{2}/30$.
 Coefficients
$C_{1-4}$ correspond to those from RK06.  After applying the mixing-length approximation to RK06's results
we get $C_{1}=-2\tau_{c}^{2}/5$, $C_{2}=-4\tau_{c}^{2}/15$,$C_{3}=0$,$C_{4}=-\tau_{c}^{2}/5$.
In confronting these coefficients to ours, we see the difference.
It can be explained, in part, by the crudeness of the given variant of tau approximation.
The comparison with RS06 is given
in Appendix B. As shown by \cite{kle-rog:07p} the spectral $\tau$ approximation is
capable to give result in closer agreement with those of SOCA.

In the commonly accepted scheme of the solar $\alpha\Omega$ dynamo, the
poloidal LSMF of the Sun is produced from the large-scale toroidal magnetic
field via the alpha effect. Expressions (\ref{dif1},\ref{difs1}) hold contributions which are capable to
induce the MEMF along the LSMF and consequently these terms are potentially very
important for the solar dynamo  because they provides additional
induction sources of the large-scale poloidal magnetic field of the Sun. 
Below, I consider the efficiency of induction effect along the nonuniform LSMF
due to global rotation and shear.

For the sake of simplicity we restrict consideration to the axisymmetric LSMF in the Keplerian
disk  in the disk geometry. In cylindrical coordinates $\left(r,\phi,z\right)$ the axisymmetric
LSMF can be expressed via
 $\mathbf{\overline{B}}=B\mathbf{e_{\phi}}+\mathbf{rot}\left(A\mathbf{e}_{\phi}\right)$
and the global rotation velocity is
$\overline{\mathbf{V}}=r\Omega\mathbf{e}_{\phi}$. We assume that
toroidal LSMF exceeds its poloidal counterpart,
$\mathbf{\overline{B}}\approx B\mathbf{e_{\phi}}$.
 In  (\ref{dif1},\ref{difs1})
we leave only those terms that induce the toroidal MEMF and skip the
usual contributions due to turbulent diffusion as well.

In the Keplerian disk we have $\partial\log\Omega/\partial\log
r=-3/2$. For the given conditions the contribution
of shear in \ref{difs1} is defined by terms at $C_3,C_4$. It is calculated
as follows 

\begin{equation}
\mathcal{E}^{(V)}_{\phi}\approx\left(
C_3(\nabla_r\overline{\mathbf{V}})_{\phi}(\nabla_z\overline{\mathbf B})_{\phi}
+C_4(\nabla_z\overline{\mathbf
  B})_{\phi}(\nabla_{\phi}\overline{V})_{r}\right)\left\langle
u^{(0)2}\right
\rangle, \label{e-cov}
\end{equation}
where covariant derivatives are
$(\nabla_r\overline{\mathbf
  V})_{\phi}=r\partial_r(\overline{V}_{\phi}/r)$,$(\nabla_{\phi}\overline{\mathbf V})_r=-\overline{V}_{\phi}/r$ and
$(\nabla_z\overline{\mathbf B})_{\phi}=\partial_z{B}$. 
Then, the contribution of shear to the MEMF is defined by
 $r\partial_r\left(\overline{V}_{\phi,r}/r\right)\tau_{c}=-3\Omega^{*}/4$ and $-\tau_c\overline{V}_{\phi}/r=-.5\Omega^{*}$.
Our derivations are valid in the case of the weak shear flow, $\left|V_{i,j}\tau_{c}\right|\ll1$.
For the Keplerian disks this condition is fulfilled if $\Omega^{*}\ll 1$.
In taking the latter into account and using (\ref{dif0}),
we find the azimuthal component of the MEMF generated from the non-uniform
toroidal component of LSMF via effects of the global rotation and shear,
 \begin{equation}
\mathbf{\mathcal{E}}_{\phi}\approx  \frac{2\Omega^{*}}{15}(2\varepsilon+1)
\langle u^{(0)2}\rangle \tau_{c}
 \frac{\partial B}{\partial z}.\label{ts1}\end{equation}
Therefore if the LSMF is concentrated to the plan of disk
the induced MEMF is in direction of the LSMF.

\subsection{Slow rotation, arbitrary LSMF }

\subsubsection{Spatially uniform LSMF}

In this part of the paper we consider results obtained for the slow
rotation limit. In what follows, no restriction is applied to the
strength of the LSMF.  The MEMF, that is induced due to influence of
rotation and stratification on the turbulence, is described with expression
\begin{eqnarray}
\mathcal{E}^{(a)} & = & \left\langle u^{(0)2}\right\rangle \tau_{c}\left\{ \varphi_{1}^{(a)}\left(\mathbf{G}\times\mathbf{\overline{B}}\right)+\varphi_{2}^{(a)}\left(\mathbf{U}\times\mathbf{\overline{B}}\right)+\tau_{c}\left(\mathbf{\Omega\cdot\overline{B}}\right)\left(\varphi_{4}^{(a)}\mathbf{G}+\varphi_{10}^{(a)}\mathbf{U}\right)\right.\label{al:nl}\\
 & + & \mathbf{\tau_{c}\overline{B}\left(\varphi_{6}^{(a)}\left(\mathbf{\Omega\cdot G}\right)+\varphi_{8}^{(a)}\left(\Omega\mathbf{\cdot U}\right)\right)}+\tau_{c}\mathbf{\Omega}\left(\varphi_{5}^{(a)}\left(\mathbf{\overline{B}\cdot G}\right)+\varphi_{9}^{(a)}\left(\mathbf{\overline{B}\cdot U}\right)\right)\nonumber \\
 & + &
 \mathbf{\left.\tau_{c}\frac{\left(\mathbf{\Omega\cdot\overline{B}}\right)\overline{B}}
{\overline{B}^{2}}\left(\varphi_{3}^{(a)}\left(\mathbf{\overline{B}\cdot
      G}\right)
+\varphi_{7}^{(a)}\left(\mathbf{\overline{B}\cdot U}\right)\right)\right\} }
+
\tau_{c}h_{\mathcal{C}}^{(0)}\varphi_{1}^{(h)}\overline{\mathbf{B}},\nonumber
\end{eqnarray}
where $\varphi_{n}^{\left(a\right)}$ are functions of $\beta$ defined
in the appendix. This formula generalizes the similar results by \citet{kit-rud:1993b,kit-rud:1993a}
taking the density stratification, magnetic fluctuations and current
helicity into account. The nonlinear MEMF of helical
MHD turbulence was considered by \citet{kle-rog:04c}, as well. For
the strong LSMF limit we obtain
 \begin{eqnarray}
\mathcal{E}^{(a)}|_{\beta\rightarrow\infty} & = & \left\{ \frac{\tau_{c}}{8}\left(\left(\varepsilon+1\right)\left(\mathbf{\overline{B}\cdot U}\right)+\frac{3\left(3\varepsilon+5\right)}{8}\left(\mathbf{\overline{B}\cdot G}\right)\right)\left(\mathbf{\Omega}-\frac{\left(\mathbf{\Omega\cdot\overline{B}}\right)\mathbf{\overline{B}}}{\overline{B}^{2}}\right)\right.\label{al:nl:sb}\\
 & + & \left.\frac{3\varepsilon+1}{64}\left(\mathbf{G}\times\mathbf{\overline{B}}\right)\right\} \frac{\pi}{\beta}\left\langle u^{(0)2}\right\rangle \tau_{c}.\nonumber \end{eqnarray}
 The results by \citet{kit-rud:1993b} can be recovered from (\ref{al:nl:sb}),
if we put $G=0$ and $\varepsilon=0$. Following to arguments given in
the paper cited above, we conclude that the MEMF like (\ref{al:nl:sb})
does not produce a dynamo. 

The first term at the upper line of (\ref{al:nl}) describes the so-called
``turbulent buoyancy'' (\citet{kit-rud:1993a}). The expression
(\ref{al:nl:sb}) shows that the transport of LSMF is downward for
the strong magnetic field limit. For the case of the weak field we
get $\varphi_{1}^{(a)}\approx\varepsilon/6+\left(6\varepsilon-8\beta^{2}\right)/15$.
Then, if we neglect contributions due to small-scale magnetic fluctuations,
we obtain that for the weak field transport is upward (opposite to
direction of $\mathbf{G}$). In this case the effective drift velocity
is proportional to the LSMF's pressure (\citet{kit-rud:1993a}). In
this aspect it is similar to the usual buoyancy of magnetic flux tubes
(\citet{park}). Furthermore, we find that the large-scale inhomogeneity
of magnetic fluctuations provide the downward drift of LSMF in the
whole range of magnetic field strength.

\begin{figure}
\begin{centering}\includegraphics[width=7cm,height=11cm,angle=-90]{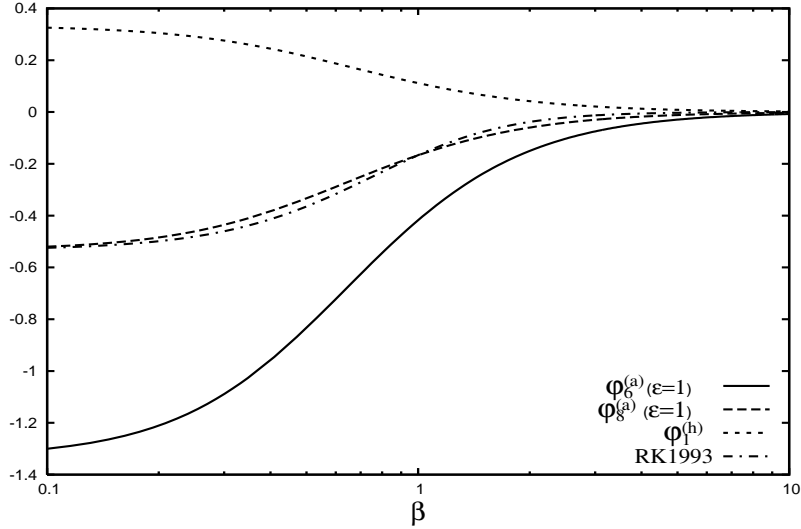}\par\end{centering}

\caption{\label{cap:a-quenching}The quenching functions for isotropic components
of $\alpha$-effect.}
\end{figure}

The quenching functions for the isotropic
components of $\alpha$ effect are shown on Fig.\ref{cap:a-quenching}.
There, for comparison, via the dash-dotted line, we show the curve
corresponding to quenching of isotropic components of $\alpha$ effect
obtained within SOCA in (\citet{kit-rud:1993b}).%

In the strong LSMF limit we found that $\alpha$ effect is quenched
as $\beta^{-2}$ which is different from results by \citet{kit-rud:1993b}
and similar to findings by \citet{kle-rog:04c}. Though, as seen from
the figure, the numerical difference between the quenching curves
obtained within SOCA (dash-dotted line) and MTA (dashed line) is within
a few percents.

The non-linear electromotive force induced by shear is expressed as
follows,
\begin{eqnarray}
\mathcal{E}_{i}^{(s)} & = & \varepsilon_{inm}\left\{
 \varphi_{1}^{(s)}\frac{\overline{B}_l\overline{B}_k}{\overline{B}^{2}}\overline{V}_{l,k}U_{n}\overline{B}_{m}
+ \varphi_{2}^{(s)}\overline{B}_l\overline{V}_{l,m}U_{n}
+\varphi_{3}^{(s)}(\mathbf{U}\cdot\mathbf{\overline{B}})\frac{\overline{B}_l\overline{B}_m}{\overline{B}^{2}}
(\overline{V}_{l,n}-\overline{V}_{n,l})\right.\label{eq:trsh}\\
&+& \left. \varphi_{4}^{(s)}U_{l}\overline{V}_{n,l}\overline{B}_m+ \varphi_{5}^{(s)}U_{n}\overline{V}_{l,m}\overline{B}_l
 \right\} \langle u^{(0)2}\rangle \tau_{c}^{2}\nonumber \\
&+& \tau_{c}^{2}h_{\mathcal{C}}^{(0)}\left\{
  \varphi_{4}^{(h)}\overline{V}_{m,n}
\frac{\overline{B}_{m}\overline{B}_{n}}{\overline{B}^{2}}\overline{B}_{i}
+\varphi_{3}^{(h)}(\overline{V}_{n,i}+\overline{V}_{i,n})\overline{B}_{n}
+\varphi_{2}^{(h)}\left(\mathbf{W}\times\mathbf{\overline{B}}\right)_i\right\}
,\nonumber 
\end{eqnarray}
From the structure of (\ref{eq:trsh}) we can conclude that  contributions with $\varphi_{2,3,5}^{(s)}$ and the second term
in brackets with $\varphi_{5}^{(5)}$ can be interpreted as the
$\alpha$ effect. The terms with $\varphi_{1,4}^{(s)}$ and
$\varphi_{2}^{(h)}$ provide the pumping of LSMF. 
Surprisingly, the $\alpha$-effect like terms survive even in
the limit of the strong magnetic field. In this case we get
\begin{eqnarray}
\mathcal{E}_{i}^{(s)}|_{\beta\rightarrow\infty} & = & \varepsilon_{inm}\left\{
\frac{3}{4}(\varepsilon-1)\left(\frac{\overline{B}_l\overline{B}_k}{\overline{B}^{2}}\overline{V}_{l,k}U_{n}\overline{B}_{m}
-\overline{B}_l\overline{V}_{l,m}U_{n}\right)\right.\label{eq:trsh:str}\\
&+&\left.(\varepsilon+1)(\mathbf{U}\cdot\mathbf{\overline{B}})\frac{\overline{B}_l\overline{B}_m}{\overline{B}^{2}}
(\overline{V}_{l,n}-\overline{V}_{n,l}) \right\}
\frac{\pi}{16\beta}\left\langle u^{(0)2}\right\rangle \tau_{c}^{2}\nonumber \\
&+&\frac{3\pi}{64\beta} \tau_{c}^{2}h_{\mathcal{C}}^{(0)}\left\{ \overline{V}_{m,n}
\frac{\overline{B}_{m}\overline{B}_{n}}{\overline{B}^{2}}\overline{B}_{i}
-(\overline{V}_{n,i}+\overline{V}_{i,n})\overline{B}_{n}
+\left(\mathbf{W}\times\mathbf{\overline{B}}\right)_i\right\},\nonumber 
\end{eqnarray}
According to (\ref{emfas1}) and (\ref{eq:trsh:str}) the pumping of the
LSMF due to joint effect of current helicity
and shear have the same sign for the weak and strong LSMF.
\begin{figure}
\begin{centering}\includegraphics[height=11cm,keepaspectratio,angle=-90]{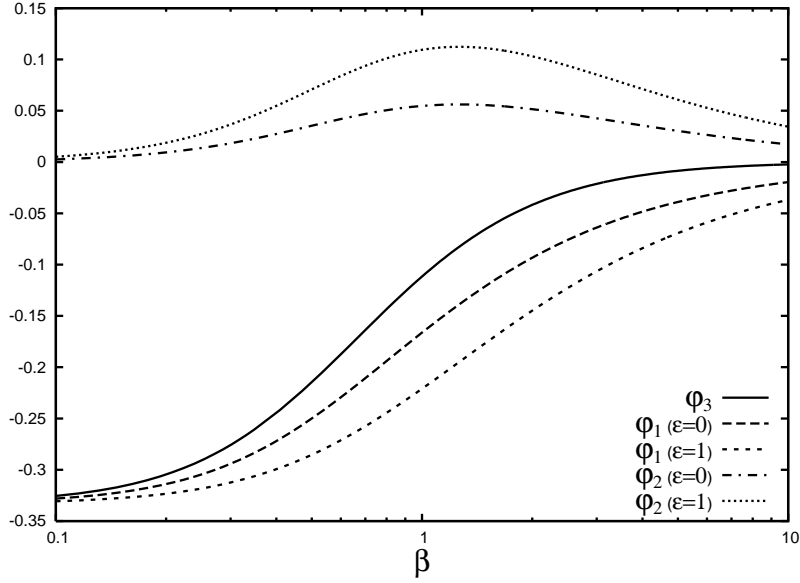}\par\end{centering}

\caption{\label{cap:eta-quench}Functions defining the nonlinear turbulent
diffusion of LSMF (see eq (\ref{eq:eta-quench})). }
\end{figure}

\subsubsection{Diffusion, $\Omega\times\mathbf{J}$ and shear current effect\label{sub:diff-shear}}

The results for nonlinear turbulent diffusion are  similar to those
found within SOCA by \citet{kit-pip-rud}. We have
 \begin{equation}
\mathcal{E}^{\left(d\right)}=\left\{ \varphi_{3}\mathbf{\nabla\times\overline{B}}+\left(\varphi_{2}\frac{\left(\left(\mathbf{\nabla\times\overline{B}}\right)\times\mathbf{\overline{B}}\right)}{\overline{B}^{2}}+\varphi_{1}\mathbf{\nabla}\log\left(\frac{\overline{B}^{2}}{2}\right)\right)\times\overline{\mathbf{B}}\right\} \left\langle u^{(0)2}\right\rangle \tau_{c}+\mathbf{\mathcal{E}}^{(w)},\label{eq:eta-quench}\end{equation}
where $\mathbf{\mathcal{E}}^{(w)}$ stands for the contributions due
to rotation. The corresponding quenching functions are given on Fig.\ref{cap:eta-quench}.

\begin{figure}
\begin{centering}\includegraphics[height=11cm,keepaspectratio,angle=-90]{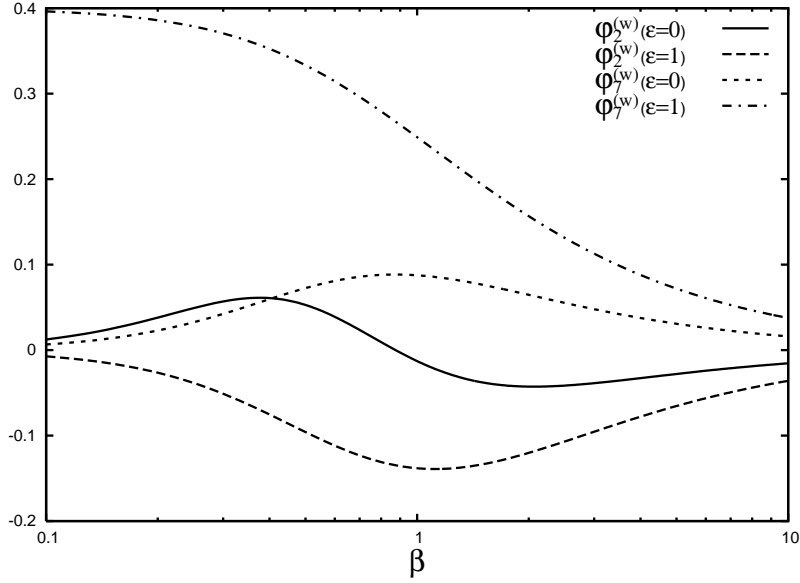}\par\end{centering}

\caption{\label{cap:omegaxj}The quenching functions for {}``$\Omega\times\mathbf{J}$''
generation effect for different parameters .}
\end{figure}

The next formula generalizes the results for the nonlinear diffusion
of LSMF to the case of the slowly rotating media,
\begin{eqnarray}
\mathcal{E}_{i}^{(w)} & = & \left\{ \varphi_{8}^{(w)}\nabla_{i}\left(\mathbf{\Omega\cdot\overline{B}}\right)+\varphi_{1}^{(w)}\frac{\left(\mathbf{\Omega\cdot\overline{B}}\right)}{2}\nabla_{i}\log\left(\overline{B}^{2}\right)+\varphi_{4}^{(w)}\overline{B}_{i}\frac{\left(\mathbf{\overline{B}}\cdot\nabla\right)\left(\mathbf{\Omega\cdot\overline{B}}\right)}{\overline{B}^{2}}\right.\nonumber \\
 & +\!\! & \varphi_{5}^{(w)}\Omega_{i}\frac{\left(\mathbf{\overline{B}}\cdot\nabla\right)}{2}\log\left(\overline{B}^{2}\right)\!\!+\varphi_{3}^{(w)}\overline{B}_{i}\frac{\left(\Omega\cdot\mathbf{\overline{B}}\right)}{\overline{B}^{2}}\frac{\left(\mathbf{\overline{B}}\cdot\nabla\right)}{2}\log\left(\overline{B}^{2}\right)\label{eq:oxj:nl}\\
 & + & \left.\varphi_{6}^{(w)}\frac{\left(\Omega\cdot\mathbf{\overline{B}}\right)}{\overline{B}^{2}}\left(\mathbf{\overline{B}}\cdot\nabla\right)\overline{B}_{i}+\varphi_{2}^{(w)}\overline{B}_{i}\frac{\left(\Omega\cdot\nabla\right)}{2}\log\left(\overline{B}^{2}\right)+\varphi_{7}^{(w)}\left(\mathbf{\Omega}\cdot\nabla\right)\overline{B}_{i}\right\} \left\langle u^{(0)2}\right\rangle \tau_{c}^{2}.\nonumber \end{eqnarray}
The last two terms at the third line in (\ref{eq:oxj:nl}) are related
with the generation of MEMF along the direction of LSMF. The corresponding
functions $\varphi_{2}^{(w)}$ and $\varphi_{7}^{(w)}$ are shown
on Fig.\ref{cap:omegaxj}. As can be seen there, in the absence of
the background magnetic fluctuations ($\varepsilon=0$) the generation
due to $\Omega\times\mathbf{J}$-effect exists only in nonlinear regime.

If  $\beta>1$, functions $\varphi_{2}^{(w)}$ and $\varphi_{7}^{(w)}$
have opposite signs everywhere. Note, while the term $\left(\mathbf{\Omega}\cdot\nabla\right)\overline{B}_{i}$
induces MEMF in direction of LSMF's gradients along axis of rotation,
the term ${\displaystyle \overline{B}_{i}\left(\Omega\cdot\nabla\right)\log\left(\overline{B}^{2}\right)}$
induces MEMF in opposite direction. Formally, the latter effect is
similar to $\alpha$-effect. The only difference with the standard
$\alpha$ is that instead stratification parameters of turbulence
we have a parameter which is related with nonuniform distribution
of the LSMF's energy. For the solar magnetic fields the effect is
antisymmetric about equator. Below, it is shown that in the strong
LSMF this $\alpha$ is quenched by factor $\beta^{_{-1}}$ which is
lesser than for standard $\alpha$. 

For the limit of the strong LSMF we get
\begin{eqnarray}
\mathcal{E}_{i}^{(w)}|_{\beta\rightarrow\infty} & = & \left\{
  \left(17\varepsilon+47\right)
\left(\nabla_{i}\left(\mathbf{\Omega\cdot\overline{B}}\right)
-\frac{\left(\mathbf{\Omega\cdot\overline{B}}\right)}{2}\nabla_{i}
\log\left(\overline{B}^{2}\right)\right)\right.\label{eq:omxj-limit}\\
 & - & \left(21\varepsilon+43\right)\left(\Omega_{i}\frac{\left(\mathbf{\overline{B}}\cdot\nabla\right)}{2}\log\left(\overline{B}^{2}\right)+\overline{B}_{i}\frac{\left(\mathbf{\overline{B}}\cdot\nabla\right)\left(\Omega\cdot\mathbf{\overline{B}}\right)}{\overline{B}^{2}}+\frac{\left(\Omega\cdot\mathbf{\overline{B}}\right)}{\overline{B}^{2}}\left(\mathbf{\overline{B}}\cdot\nabla\right)\overline{B}_{i}\right)\nonumber \\
 & + & 3\left(21\varepsilon+43\right)\frac{\left(\Omega\cdot\mathbf{\overline{B}}\right)\left(\mathbf{\overline{B}}\cdot\nabla\right)}{2\overline{B}^{2}}\log\left(\overline{B}^{2}\right)\overline{B}_{i}\nonumber \\
 & - & \left.\left(37\varepsilon+27\right)\left(\overline{B}_{i}\frac{\left(\Omega\cdot\nabla\right)}{2}\log\left(\overline{B}^{2}\right)-\left(\mathbf{\Omega}\cdot\nabla\right)\overline{B}_{i}\right)\right\} \frac{\pi}{512\beta}\left\langle u^{(0)2}\right\rangle \tau_{c}^{2}.\nonumber \end{eqnarray}
From there we find that $\Omega\times\mathbf{J}$-effect maintain
the generation part of the MEMF even for the strong LSMF. The amplitude
of effect tends to constant as the strength of LSMF is increased.
It hardly possible to make a definite conclusion about the dynamo
effect in this case because the generation part of (\ref{eq:omxj-limit})
is contributed by terms with opposite signs.

The MEMF's contributions due to shear are defined by,

\begin{eqnarray}
\mathcal{E}_{i}^{(V)} & = & \left\langle u^{(0)2}\right\rangle
\tau_{c}^{2}\varepsilon_{inm}\left\{
\varphi_{1}^{(V)}\overline{V}_{n,l}\overline{B}_{l,m}
+\frac{\overline{B}_{l}\overline{B}_{m}}{\overline{B}^{2}}\overline{V}_{l,k}
(\varphi_{2}^{(V)}\overline{B}_{k,n}+\varphi_{3}^{(V)}\overline{B}_{k,m})
\right.
\label{eq:nlshear}\\
& + & 
(\varphi_{4}^{(V)}\overline{V}_{n,l}+\varphi_{5}^{(V)}\overline{V}_{l,n})\overline{B}_{m,l}
+\varphi_{6}^{(V)}\frac{\overline{B}_{l}\overline{B}_{n}}{\overline{B}^{2}}\overline{V}_{l,k}\overline{B}_{m,k}
+\varphi_{7}^{(V)}\frac{\overline{B}_{k}\overline{B}_{l}}{\overline{B}^{2}}\overline{V}_{l,k}\overline{B}_{m,n}
\nonumber \\
 & + & \left.
\varphi_{8}^{(V)}\frac{\overline{B}_{k}\overline{B}_{l}}{\overline{B}^{2}}\overline{V}_{m,n}\overline{B}_{l,k}
+\varphi_{9}^{(V)}\overline{V}_{l,n}\overline{B}_{l,m}
+\frac{\overline{B}_{k}\overline{B}_{l}}{\overline{B}^{2}}\overline{V}_{l,n}
(\varphi_{10}^{(V)}\overline{B}_{k,m}+\varphi_{11}^{(V)}\overline{B}_{m,k})
\right\} ,\nonumber \end{eqnarray}
where, for the sake of simplicity, we leave only the largest contributions
and those which are important for the solar-type dynamo models, where
the strength of LSMF component along direction of the large-scale flow 
dominates components directed along the shear. Reader can find the expressions
for $\varphi_{n}^{(V)}$ in Appendix A. The full expression has a much more complicated tensorial structure
than (\ref{eq:nlshear}). In the case of the strong LSMF we get,
\begin{eqnarray}
\mathcal{E}_{i}^{(V)}|_{\beta\rightarrow\infty} & = & \frac{\tau_{c}^{2}}{6}\left\langle u^{(0)2}\right\rangle
\varepsilon_{inm}\left\{
\left(\frac{\varepsilon+15}{20}\overline{V}_{n,l}-
\frac{\varepsilon}{5}\overline{V}_{l,n}\right)\overline{B}_{m,l}
+(\varepsilon+1)\frac{\overline{B}_{l}\overline{B}_{n}}{\overline{B}^{2}}\overline{V}_{l,k}\overline{B}_{m,k}
\right.
\label{eq:nlshear-lim}\\
& - & 
\frac{3\varepsilon+13}{20}\overline{V}_{n,l}\overline{B}_{l,m}
-\frac{\varepsilon+3}{2}\frac{\overline{B}_{l}\overline{B}_{m}}{\overline{B}^{2}}\overline{V}_{l,k}
\overline{B}_{k,m}+\frac{\varepsilon+1}{10}\overline{V}_{l,n}\overline{B}_{l,m}
\nonumber \\
 & + & \left.
\frac{\overline{B}_{k}\overline{B}_{l}}{\overline{B}^{2}}\overline{V}_{l,n}
\left((\varepsilon+1)\overline{B}_{m,k}-\frac{\varepsilon+3}{4}\overline{B}_{k,m}\right)
\right\} ,\nonumber \end{eqnarray}

\begin{figure}
\begin{centering}\includegraphics[height=11cm,keepaspectratio,angle=-90]{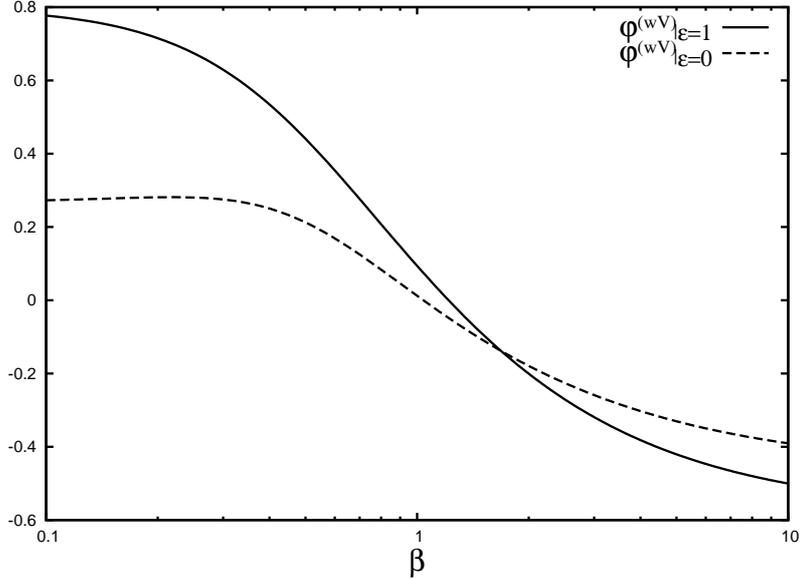}\par\end{centering}

\caption{\label{cap:oxjnl}The dependence of induction effect along the
  nonuniform toroidal LSMF  on the strength of magnetic field.}
\end{figure}

Now, we would like to consider  efficiency of induction effect along the nonuniform LSMF
due to global rotation and shear in nonlinear
regimes for the Keplerian discs. As before, we assume a disc penetrated
by the large-scale toroidal magnetic field that is nonuniform along
the axis of rotation. From (\ref{eq:omxj-limit},\ref{eq:nlshear})
we get
\begin{equation}
\mathbf{\mathcal{E}}_{\phi}\approx\frac{\Omega^{*}}{2}\left\langle u^{(0)2}\right\rangle
\tau_{c}\varphi^{(wV)}{\displaystyle \frac{\partial B}{\partial z}}.\label{eq:oxjnl}\end{equation}
where  the quenching function is $\varphi^{(wV)}=\varphi_{2}^{(w)}+\varphi_{7}^{(w)}+1.5(\varphi_{9}^{(V)}+\varphi_{10}^{(V)})+\varphi_{1}^{(V)}$.
Note, eq.(\ref{eq:oxjnl}) transforms to eq.(\ref{ts1}) in limit
$\beta\rightarrow 0$. The dependence of $\varphi^{(wV)}$ on the LSMF's strength is shown on the Fig.\ref{cap:oxjnl}.%

Results given on the Fig.\ref{cap:oxjnl} show that the $\varphi^{(wV)}$
is positive for $\beta<1$ and negative for $\beta>1$ for all
$\varepsilon$. This result
supports an idea about the change of dynamo type in passing from linear
to non-linear regime of the LSMF's generation by $\Omega\times\mathbf{J}$
and shear-current effects. Previously we found that the induction term due to
$\Omega\times\mathbf{J}$ effect tends to constant when
$\beta\rightarrow \infty$ (eq.(\ref{eq:omxj-limit})) while the
induction term due
to shear-current effect is growing under $\beta\rightarrow \infty$
(eq. (\ref{eq:nlshear-lim})). Therefore, the primary nonlinear
generation effect in the differentially
rotating \emph{uniform} MHD turbulence penetrated by the  \emph{nonuniform}
toroidal LSMF may be due to shear-current effect.  The same nonlinear
dependence of shear-current effect was discovered in the paper by \citet{kle-rog:04c} for the different
kind of MTA. In the next section I show that the given sources of the
MEMF  ultimately result to current helicity generation. Therefore,
the effect considered above is saturated dynamically due to magnetic
helicity conservation law.

\section{The current helicity evolution\label{sec:The-current-helicity}}

As we have seen, the current helicity contributes to the different
kind of MEMF's action, not only to the $\alpha$ effect. The recent
papers \citep{sub-bra:04} show that the magnetic helicity conservation
law can be described in terms of the current helicity evolution if
the assumption of the scale separation is fulfilled. For the time
being the redistribution of current helicity over the space scales
is not satisfactory understood. One attempt to describe the helicity
evolution in turbulent media penetrated by LSMF was given in the papers
by \citet{bra-sub:04,sub-bra:04}. Here, we will follow their results
and obtain the explicit evolutionary equation for the current helicity.
The equation in question can be derived from (\ref{induc2},\ref{navie2}).
After integration over the large-scale variables we can get the general
equation for the current helicity in the following form,
\begin{eqnarray}
\frac{\partial h_{\mathcal{C}}}{\partial t} & =\!\! &
-\frac{h_{\mathcal{C}}}{\tau_{h}}
+\frac{2}{\mu\rho}\varepsilon_{plm}\int\left[k^{2}\hat{\varkappa}_{lp}\frac{\overline{B}_{m}}{\rho}
-\mathrm{i}\hat{\varkappa}_{lp}\left(\mathbf{k}\cdot\nabla\right)\left(\frac{\overline{B}_{m}}{\rho}\right)
-\frac{\mathrm{i}}{2}\left(\mathbf{k\cdot\nabla}\right)\left(\hat{\varkappa}_{lp}
\frac{\overline{B}_{m}}{\rho}\right)\right.\label{curhel0}\\
 & +\!\! &
 \left.\mathrm{i}k_{p}\nabla_{n}\left(\hat{\varkappa}_{ln}\frac{\overline{B}_{m}}{\rho}\right)
+\frac{1}{2}\overline{V}_{l,n}\left(\mathrm{i}k_{p}-\frac{1}{2}\nabla_{p}\right)\left(\hat{h}_{mn}
-\hat{h}_{nm}\right)-\frac{1}{2}\overline{V}_{l,m}\nabla_{n}\hat{h}_{np}\right]\mathd\mathbf{k}.\nonumber \end{eqnarray}
The third order moments were replaced by $-h_{\mathcal{C}}/\tau_{h}$
, $\tau_{h}$- is a relaxation time for the current helicity . This
is a rather rough way because the triple correlations may give important
contribution for the helicity redistribution over the space scales
\citep{pouquet-al:1975a,kleruz82,kle-rog99}. Because of the very
rough assumptions used in derivation of (\ref{curhel0}), it should
be considered with caution. In spite of the latter, the equation (\ref{curhel0})
provides a useful tool for investigation the nonlinear saturation
in helical mean-field dynamo\citep{bra-sub:04d}. Except for contributions
due to density stratification and shear, equation (\ref{curhel0})
can be reproduced from results of BS05 after substitution identity
$\varepsilon_{ijk}\varepsilon_{ipq}\varepsilon_{qlm}=\varepsilon_{lmk}\delta_{jp}-\varepsilon_{lmj}\delta_{kp}$
in eq. (10.71) there. Inspection of (\ref{curhel0}) shows that if
we replace $k^{2}\rightarrow\ell_{c}^{-2}$ and use (\ref{cor2}),
we can write the evolutionary equation in the following form,
 \begin{eqnarray}
\frac{\partial h_{\mathcal{C}}}{\partial t} & =\!\! &
-\frac{2\left(\mathbf{\mathcal{E}}\cdot\mathbf{\overline{B}}\right)}{\mu\rho\ell_{c}^{2}}
-\frac{h_{\mathcal{C}}}{\tau_{h}}+\frac{2}{\mu\rho}\varepsilon_{plm}\int\left[-\mathrm{i}\hat{\varkappa}_{lp}k^{n}\nabla_{n}\left(\frac{\overline{B}_{m}}{\rho}\right)
-\frac{\mathrm{i}}{2}\left(\mathbf{k\cdot\nabla}\right)\left(\hat{\varkappa}_{lp}\frac{\overline{B}_{m}}{\rho}\right)\right.\label{curhel01}\\
 & +\!\! &
 \left.\mathrm{i}k_{p}\nabla_{n}\left(\hat{\varkappa}_{ln}\frac{\overline{B}_{m}}{\rho}\right)
+\frac{1}{2}\overline{V}_{l,n}\left(\mathrm{i}k_{p}
-\frac{1}{2}\nabla_{p}\right)\left(\hat{h}_{mn}-\hat{h}_{nm}\right)
-\frac{1}{2}\overline{V}_{l,m}\nabla_{n}\hat{h}_{np}\right]\mathd\mathbf{k}.
\nonumber \end{eqnarray}
According to \citep{pouquet-al:1975a,kleruz82,vain:83,bran:01,vish-ch:01}
the first term in (\ref{curhel01}) is responsible for helicity generation
in turbulent medium. The rest part of equation can be interpreted
as the helicity fluxes \citep{vish-ch:01,sub-bra:04,sub-bra:05}.
The given expression for helicity fluxes is incomplete because the
contribution of the third order moments is dropped in (\ref{curhel01}).
As the first step we consider the case of the weak LSMF. From (\ref{curhel01})
and (\ref{secm2},\ref{eq:secm2a},\ref{eq:secm2b}) we get
\begin{eqnarray}
\frac{\partial h_{\mathcal{C}}}{\partial t}+\frac{1}{\tau_{h}}h_{\mathcal{C}}\!\! & = & -\frac{2}{\mu\rho\ell_{c}^{2}}\left(\mathbf{\mathcal{E}}\cdot\mathbf{\overline{B}}\right)+\frac{\left(\varepsilon-1\right)}{\mu\rho\tau_{c}}\left\{ 2f_{1}^{(a)}\left(\mathbf{e}\cdot\overline{\mathbf{B}}\right)\left(\mathbf{e}\cdot\left(\mathbf{U\times\overline{\mathbf{B}}}\right)\right)\right.\label{eq:hc-evom}\\
 & + & \frac{\left(\mathbf{e}\mathbf{\cdot G}\right)}{3}\left(f_{4}^{(d)}\overline{B}^{2}+f_{3}^{(d)}\left(\mathbf{e}\cdot\overline{\mathbf{B}}\right)^{2}\right)+2f_{2}^{(a)}\left(\overline{\mathbf{B}}\cdot\left(\nabla\times\overline{\mathbf{B}}\right)\right)\nonumber \\
 & + & \left(\mathbf{e}\cdot\overline{\mathbf{B}}\right)\left(\frac{1}{3}f_{4}^{(d)}\left(\mathbf{\overline{B}\cdot G}\right)+\frac{4f_{9}^{(a)}}{\left(\varepsilon+1\right)}\left(\mathbf{\overline{B}\cdot U}\right)\right)-f_{4}^{(d)}\frac{\left(\mathbf{e}\cdot\nabla\right)}{6}\overline{B}^{2}\nonumber \\
 & - & \left.\frac{4}{3}f_{1}^{(a)}\left(\mathbf{e}\cdot\overline{\mathbf{B}}\right)\left(\mathbf{e}\cdot\mathbf{\left(\nabla\times\overline{\mathbf{B}}\right)}\right)-f_{3}^{(d)}\frac{\left(\mathbf{e\cdot\nabla}\right)}{6}\left(\mathbf{e}\cdot\mathbf{\overline{B}}\right)^{2}-f_{4}^{(d)}\frac{\left(\mathbf{B}\cdot\nabla\right)}{3}\left(\mathbf{e}\cdot\mathbf{\overline{B}}\right)\right\} ,\nonumber \end{eqnarray}
where substitution $\left\langle u^{(0)2}\right\rangle \ell_{c}^{-2}\rightarrow\tau_{c}^{-2}$
was used, and $\mathbf{e}=\mathbf{\mathbf{\Omega}}/|\Omega|$ . Here,
we dropped the contributions due to shear because their effect to
the mean electromotive force was computed only to the zero order terms
about angular velocity. Furthermore, in (\ref{eq:hc-evom}) we kept
only those contributions which could be the most interesting from
the stellar dynamo applications standpoint. Note, for the equipartition
case, $\varepsilon=1$, helicity evolution satisfies the simple
equation:
\begin{eqnarray}
\frac{\partial h_{\mathcal{C}}}{\partial t}+\frac{1}{\tau_{h}}h_{\mathcal{C}} & = & -\frac{2\left(\mathbf{\mathcal{E}}\cdot\mathbf{\overline{B}}\right)}{\mu\rho\ell_{c}^{2}}.\label{eq:hc-evom1}\end{eqnarray}
It is in accordance with equation for the magnetic helicity density
obtained by \citet{sub-bra:05}. As an example of application of (\ref{eq:hc-evom1})
to the problem of the nonlinear saturation of alpha-effect, consider
the $\alpha^{2}$ dynamo in the fast rotation limit. For the sake
of simplicity we restrict ourselves only with the isotropic components
of $\alpha$-effect and neglect the helicity loss due to $h_{\mathcal{C}}/\tau_{h}$.
From (\ref{eq:hc-evom1}) and (\ref{frot}) we get
 \begin{equation}
\frac{\partial h_{\mathcal{C}}}{\partial t}=\frac{\pi\beta^{2}}{4\tau_{c}}\left(2\left\langle u^{(0)2}\right\rangle \Omega^{*}\left(\mathbf{e\cdot G}\right)-h_{\mathcal{C}}\right),\label{hc-evomfr}\end{equation}
where we keep the contributions of order $\Omega^{*\,-1}$ for the
current helicity, and drop the terms which are due to nonuniform LSMF.
If $L$ is the typical spatial scale of the LSMF then the eq. (\ref{hc-evomfr})
is justified when $LG\Omega^{*}\gg1$ and $\mu\rho\left|h_{\mathcal{C}}\right|\gg\left|\overline{\mathbf{B}}\cdot\left(\mathbf{\nabla\times\overline{B}}\right)\right|$.
The point to note that in (\ref{hc-evomfr}) we implicitly assume
that $h_{\mathcal{C}}^{\left(0\right)}\equiv h_{\mathcal{C}}$. It
is a shortcoming of the theory. However, this procedure is widely
used in the literature \citep{kleruz82,vain-kit:83,vish-ch:01,kleetal03,bra-sub:04d}.
With initial condition, $t=0,$ $h_{\mathcal{C}}=0,$ we write, similar
to \citet{vain:83}, the solution of eq (\ref{hc-evomfr}) as follows,
\begin{equation}
h_{\mathcal{C}}=2\Omega^{*}\left\langle u^{(0)2}\right\rangle \left(\mathbf{e\cdot G}\right)\left(1-\exp\left({\displaystyle -\frac{\pi}{4\tau_{c}}\int_{0}^{t}\beta^{2}dt}\right)\right).\label{hc-evomsol}\end{equation}
The given solution shows that under $t\rightarrow\infty$ we get $h_{\mathcal{C}}\rightarrow2\Omega^{*}\left\langle u^{(0)2}\right\rangle \left(\mathbf{e\cdot G}\right)\tau_{c}$.
On this basis, and in taking into account (\ref{frot}), we can conclude
that $\alpha$-effect will saturates exponentially under the increase
of the LSMF strength. Furthermore, this conclusion was confirmed with
numerical dynamo model which is considered by author in the separate
paper \citep{pip06a}. 

Next, we consider the equation for the current helicity evolution
for the slow rotation limit. No restriction is applied to the strength
of LSMF. The contribution of shear to the transport and generation
part of equation is described with a quite bulky tensor expressions
and we decide to restrict ourselves with terms which have either a
finite limit under $\beta\rightarrow0$ or the amplitude functions
that are greater than $0.1$. We write the evolutionary equation for
the current helicity as follows:
\begin{eqnarray}
\frac{\partial h_{\mathcal{C}}}{\partial t}+\frac{1}{\tau_{h}}h_{\mathcal{C}}\!\! & = & -\frac{2}{\mu\rho\ell_{c}^{2}}\left(\mathbf{\mathcal{E}}\cdot\mathbf{\overline{B}}\right)+\psi_{1}\frac{\overline{B}_{m}\overline{B}_{p}}{\overline{B}^{2}}\overline{V}_{p,m}h_{\mathcal{C}}+\left(\psi_{2}\mathbf{G}+\psi_{3}\mathbf{U}\right)\cdot\mathbf{W}\left\langle u^{(0)2}\right\rangle \label{eq:hc-ev}\\
 & + & \frac{1}{\mu\rho}\nabla\cdot\left(\left[\psi_{5}\nabla\times\overline{\mathbf{B}}+\psi_{4}\left(\mathbf{U\times\overline{B}}\right)\right]\left(\overline{\mathbf{B}}\cdot\overline{\mathbf{V}}\right)+\psi_{6}\mathbf{W}\overline{B}^{2}\right)+\left(\varepsilon-1\right)\left\{ ...\right\} ,\nonumber \end{eqnarray}
where $\mathbf{W}=\nabla\times\overline{\mathbf{V}}$ . Quenching
functions $\psi_{\left\{ n\right\} }$ are given in Appendix A. Symbol
$\left\{ ...\right\} $ denotes those terms which are not important
in the case $\varepsilon=1$. Taking the Taylor expansion of (\ref{eq:hc-ev})
for the case $\overline{B}\rightarrow0$ (keeping $\overline{B}^{2}$
terms) we get
 \begin{eqnarray}
\frac{\partial h_{\mathcal{C}}}{\partial
  t}+\frac{1}{\tau_{h}}h_{\mathcal{C}}\!\! & = &
-\frac{2}{\mu\rho\ell_{c}^{2}}\left(\mathbf{\mathcal{E}}\cdot\mathbf{\overline{B}}\right)-\frac{4}{15}\frac{\overline{B}_{m}\overline{B}_{p}}{\mu\rho\left\langle
    u^{(0)2}\right\rangle }V_{p,m}h_{\mathcal{C}}
-\frac{\left(\mathbf{G\cdot W}\right)}{6}\left\langle
  u^{(0)2}\right\rangle 
-\nabla\cdot\mathcal{\mathbf{\mathcal{F}}}\label{eq:hc-ev1}\\
\mathcal{\mathbf{\mathcal{F}}} & = & \left(\frac{1}{6}\left\langle
    u^{(0)2}\right\rangle 
+\frac{2}{15}\frac{\overline{B}^{2}}{\mu\rho}\right)\mathbf{W}
+\frac{2}{15\mu\rho}\left(\left[\nabla\times\overline{\mathbf{B}}-\left(\mathbf{U\times\overline{B}}\right)\right]\left(\overline{\mathbf{B}}\cdot\overline{\mathbf{V}}\right)\right),\label{eq:helF}\end{eqnarray}
where we apply the equipartition condition, $\varepsilon=1$, as well.
The direction of the helicity flux due to the first contribution in
(\ref{eq:helF}), $\mathbf{\mathcal{F}}^{\mathbf{W}}={\displaystyle \left(\left\langle u^{(0)2}\right\rangle /6+2\overline{B}^{2}/\left(15\mu\rho\right)\right)\mathbf{W}}$,
depends on distribution of the large-scale vorticity solely. The second
term depends on details of the dynamo action. To estimate the direction
of the helicity transport due to $\mathbf{\mathcal{F}}^{\mathbf{W}}$
on the Sun we compute the vector field of the large-scale vorticity
$\mathbf{W}$. In the spherical coordinate system we have $\mathbf{W}=\sin\theta\mathbf{e}^{r}\partial\Omega/\partial\theta-r\sin\theta\mathbf{e}^{\theta}\partial\Omega/\partial r$,
where $r,\theta$ are the radial distance and the polar angle, respectively.
The distribution of the angular velocity is taken as an analytical
fit given by \citet{belved00}. It is shown at the left side Fig.\ref{fig:last}
The computed vector field of the large-scale vorticity is shown at
the right side Fig.\ref{fig:last}. 

\begin{center}%
\begin{figure}
\begin{centering}\includegraphics[width=6cm,keepaspectratio]{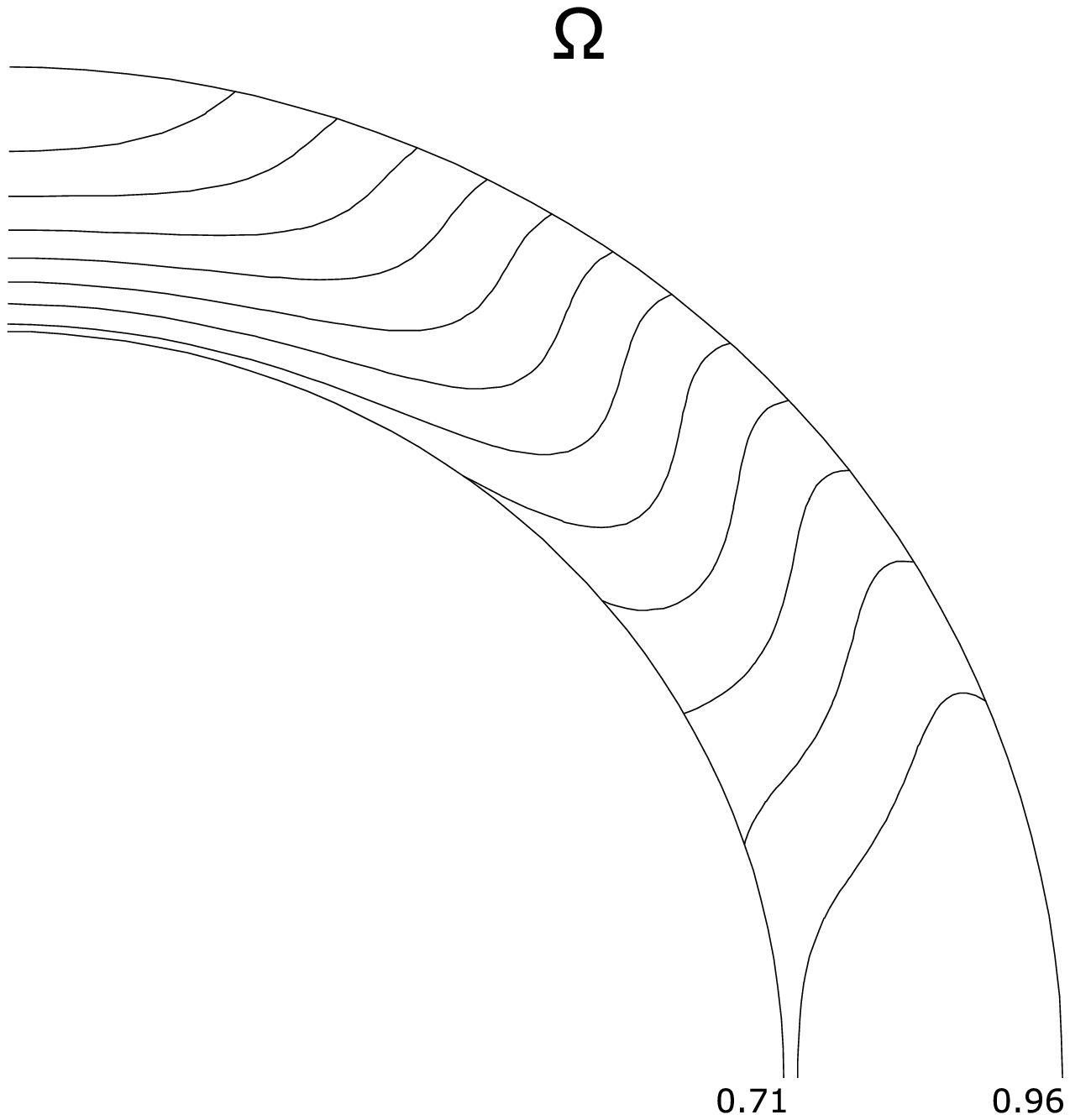}~\includegraphics[width=6cm,keepaspectratio]{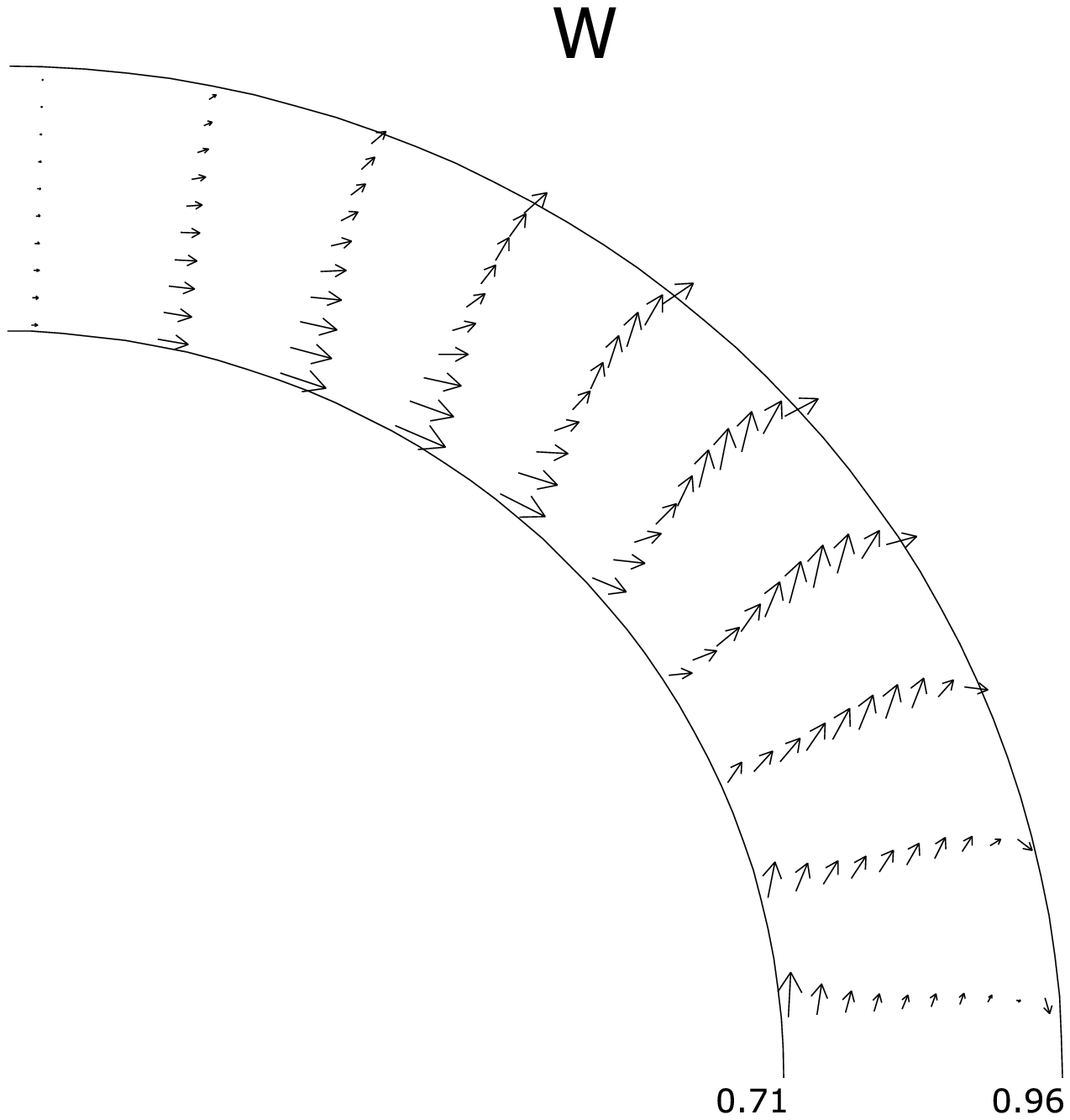}\par\end{centering}

\caption{\label{fig:last}The isolines of the angular velocity distribution
(left) and the corresponding vector field of the large-scale vorticity(right).}
\end{figure}
\par\end{center}

The given figure shows the possibility of the outward helicity flux
from the dynamo region due to shear. Note that one component of the
helicity flux $\mathbf{\mathcal{F}^{W}}$ is due to the small-scale
dynamo, ${\displaystyle \left\langle u^{(0)2}\right\rangle \mathbf{W}}/6$,
and another is due to the LSMF, $2\overline{B}^{2}\mathbf{W}/\left(15\mu\rho\right)$.
Among two the contribution of the small-scale dynamo is likely to
be dominated in the depth of convection zone. While the flux due to
the LSMF may be important at near the surface level. The latter effect
may produce the significant outward flux of helicity only with the
open boundaries \citep{bra-sub:04d,sub-bra:05}. At the near surface
level the amplitude of the large-scale vorticity, $\left|\mathbf{W}\right|\approx4\times10^{-8}s^{-1}\approx1.5\times10^{-5}day^{-1}$.
The magnitude of the surface magnetic flux change during the solar
cycle is about $10^{24}Mx$ \citep{schr-harv84}. Then the magnitude
of the helicity outflow from $2\overline{B}^{2}\mathbf{W}/\left(15\mu\rho\right)$
is about $10^{43}Mx^{2}day^{-1}$. It is compatible with estimations
given by \citet{devore:2000}. We have estimated only one part of
the helicity flux. The numerical dynamo model based on the given results
would help to get a more definite conclusions about this subject.

\section{Summary}

In the paper the mean electromotive force of turbulent flows and magnetic
fields is computed analytically using the framework of mean-field
dynamo theory and MTA (minimal $\tau$ approximation). There is an
overlap in  results obtained
with SOCA and with MTA approximation. The two approximations give qualitatively
the same results about nonlinear dependence of mean electromotive
force on the strength of LSMF or on the Coriolis number. Also there is
a difference between predictions of  SOCA and  MTA for
mean-electromotive force expressions if the shear is taken into
account. This difference 
 can be explained, in part, by the crudeness of the given version
of tau approximation.  The deficient accuracy of calculations of shear contribution
is due to an assumption about the scale-independent $\tau$. In
whole, the accuracy of the theory presented in the paper is comparable
with the mixing-length approximation. It has no the firm grounds and
should be considered with caution.

Finally, I would like to focus on the new findings of the paper. In
this study it is shown that the new interesting component of transport
of LSMF appears due to joint contribution of current helicity and
shear. The effect does not disappear in the strong LSMF limit, $\beta\gg1$.
It may be important near the base of solar CZ where the influence
of rotation and shear on the turbulence is quite strong. Furthermore,
the analysis, which we carried out for the current helicity evolution,
suggests that the shear and rotation may redistribute the helicity
in solar CZ amplifying it (in amplitude) at the near equatorial regions
in agreement with observations. Beside, the effect of rotation and
stratification on the $h_{\mathcal{C}}$ evolution is calculated explicitly.
Basically, the equation for current helicity is obtained using the
same approach as for the mean electromotive force and on the base
of quantities which are explicitly gauge invariants. Therefore, we
can expect that the dynamo model based on the above approach could
be capable for meeting the requirements of both solar and stellar
dynamo simulations.

\paragraph{Acknowledgments}

This work benefited from the  research grants 05-02-16326
and 4741.2006.2.
I thank sincerely K.-H. R\"adler and A. Brandenburg for criticism
and helpful suggestions. I acknowledge  valuable discussion with
Nathan Kleeorin and Igor Rogachevskii.
Gratitude also goes to Kirill Kuzanyan for the critical reading
manuscript. The large portion of calculations presented in
the paper were done with help of Maxima computer algebra system
(http://Maxima.sf.net). I owe sincere thanks  to all developers of Maxima and
personally to Victor Toth \citep{toth:05} for maintaining the Maxima's
package for tensor calculations.

\bibliographystyle{elsart-harv} %{plainnat}
\bibliography{dyn}

\subsection*{Appendix A.}

This part of appendix gives the functions of the Coriolis number defining
the dependence of the turbulent transport generation and diffusivities
on the angular velocity.
\begin{eqnarray*}
f_{1}^{(a)} & = & \frac{1}{4\Omega^{*\,2}}\left(\left(\Omega^{*\,2}+3\right)\frac{\arctan\Omega^{*}}{\Omega^{*}}-3\right),\\
f_{2}^{(a)} & = & \frac{1}{4\Omega^{*\,2}}\left(\left(\Omega^{*\,2}+1\right)\frac{\arctan\Omega^{*}}{\Omega^{*}}-1\right),\\
f_{3}^{(a)} & = & \frac{1}{4\Omega^{*\,2}}\left(\left(\left(\varepsilon-1\right)\Omega^{*\,2}+\varepsilon-3\right)\frac{\arctan\Omega^{*}}{\Omega^{*}}+3-\varepsilon\right),\\
f_{4}^{(a)} & = & \frac{1}{6\Omega^{*\,3}}\left(3\left(\Omega^{*4}+6\varepsilon\Omega^{*2}+10\varepsilon-5\right)\frac{\arctan\Omega^{*}}{\Omega^{*}}-\left((8\varepsilon+5)\Omega^{*2}+30\varepsilon-15\right)\right),\\
f_{5}^{(a)} & = & \frac{1}{3\Omega^{*\,3}}\left(3\left(\Omega^{*4}+3\varepsilon\Omega^{*2}+5(\varepsilon-1)\right)\frac{\arctan\Omega^{*}}{\Omega^{*}}-\left((4\varepsilon+5)\Omega^{*2}+15(\varepsilon-1)\right)\right),\\
f_{6}^{(a)} & = & -\frac{1}{48\Omega^{*\,3}}\left(3\left(\left(3\varepsilon-11\right)\Omega^{*2}+5\varepsilon-21\right)\frac{\arctan\Omega^{*}}{\Omega^{*}}-\left(4\left(\varepsilon-3\right)\Omega^{*2}+15\varepsilon-63\right)\right),\\
f_{7}^{(a)} & = & \frac{1}{48\Omega^{*\,3}}\left(3\left(\left(5\varepsilon+3\right)\Omega^{*2}+11\varepsilon+5\right)\frac{\arctan\Omega^{*}}{\Omega^{*}}-\left(4\left(\varepsilon+1\right)\Omega^{*2}+33\varepsilon+15\right)\right),\\
f_{8}^{(a)} & = & -\frac{1}{12\Omega^{*\,3}}\left(3\left(\left(3\varepsilon+1\right)\Omega^{*2}+4\varepsilon-2\right)\frac{\arctan\Omega^{*}}{\Omega^{*}}-\left(5\left(\varepsilon+1\right)\Omega^{*2}+12\varepsilon-6\right)\right),\\
f_{9}^{(a)} & = & \frac{\left(\varepsilon+1\right)}{4\Omega^{*}}\left(\frac{\arctan\Omega^{*}}{\Omega^{*}}-1\right),\\
f_{10}^{(a)} & = & -\frac{1}{3\Omega^{*\,3}}\left(3\left(\Omega^{*2}+1\right)\left(\Omega^{*2}+\varepsilon-1\right)\frac{\arctan\Omega^{*}}{\Omega^{*}}-\left(\left(2\varepsilon+1\right)\Omega^{*2}+3\varepsilon-3\right)\right),\\
f_{11}^{(a)} & = & -\frac{1}{6\Omega^{*\,3}}\left(3\left(\Omega^{*2}+1\right)\left(\Omega^{*2}+2\varepsilon-1\right)\frac{\arctan\Omega^{*}}{\Omega^{*}}-\left(\left(4\varepsilon+1\right)\Omega^{*2}+6\varepsilon-3\right)\right).\end{eqnarray*}

The dependence of turbulent diffusivities on the Coriolis number (eq.
(\ref{dif1})) is given by

\begin{eqnarray*}
f_{1}^{(d)} & = & \frac{1}{2\Omega^{*\,3}}\left(\left(\varepsilon+1\right)\Omega^{*\,2}+3\varepsilon-\left(\left(2\varepsilon+1\right)\Omega^{*\,2}+3\varepsilon\right)\frac{\arctan\left(\Omega^{*}\right)}{\Omega^{*}}\right),\\
f_{2}^{(d)} & = & \frac{1}{4\Omega^{*\,2}}\left(\left(\left(\varepsilon-1\right)\Omega^{*\,2}+3\varepsilon+1\right)\frac{\arctan\left(\Omega^{*}\right)}{\Omega^{*}}-\left(3\varepsilon+1\right)\right),\\
f_{3}^{(d)} & = & \frac{1}{2\Omega^{*\,3}}\left(3\left(3\Omega^{*\,2}+5\right)\frac{\arctan\left(\Omega^{*}\right)}{\Omega^{*}}-\left(4\Omega^{*\,2}+15\right)\right),\\
f_{4}^{(d)} & = & \frac{1}{2\Omega^{*\,3}}\left(\left(2\Omega^{*\,2}+3\right)-3\left(\Omega^{*\,2}+1\right)\frac{\arctan\left(\Omega^{*}\right)}{\Omega^{*}}\right).\end{eqnarray*}

The magnetic quenching functions of the generation and transport effects
in eq. (\ref{al:nl}) are

\begin{eqnarray*}
\varphi_{1}^{(a)} & = & \frac{1}{64\beta^{2}}\left(\left(4\left(3\varepsilon+1\right)\beta^{2}-17\varepsilon+21\right)\frac{\arctan\left(2\beta\right)}{2\beta}+\frac{\left(4\left(11\varepsilon-15\right)\beta^{2}+17\varepsilon-21\right)}{\left(4\beta^{2}+1\right)}\right),\\
\varphi_{2}^{(a)} & = & \frac{\left(1-\varepsilon\right)}{8\beta^{2}}\left(\frac{\arctan\left(2\beta\right)}{2\beta}-1\right),\\
\varphi_{3}^{(a)} & = & \frac{1}{3072\beta^{4}}\left(\frac{8\beta^{2}\left(2\beta^{2}\left(16\left(45\varepsilon+107\right)\beta^{2}-1731\varepsilon+739\right)-2097\varepsilon+97\right)-2295\varepsilon-105}{\left(4\beta^{2}+1\right)^{2}}\right.\\
 & - & \left.3\left(12\beta^{2}\left(16\left(3\varepsilon+5\right)\beta^{2}-41\varepsilon+41\right)-765\varepsilon-35\right)\frac{\arctan\left(2\beta\right)}{2\beta}\right),\\
\varphi_{4}^{(a)} & = & \frac{1}{3072\beta^{4}}\left(\frac{8\beta^{2}\left(2\beta^{2}\left(128\left(3\varepsilon+1\right)\beta^{2}+807\varepsilon-71\right)+555\varepsilon-11\right)+459\varepsilon+21}{\left(4\beta^{2}+1\right)^{2}}\right.\\
 & - & \left.3\left(4\left(115\varepsilon-19\right)\beta^{2}+153\varepsilon+7\right)\frac{\arctan\left(2\beta\right)}{2\beta}\right),\end{eqnarray*}

\begin{align*}
\varphi_{5}^{(a)} & =\frac{1}{3072\beta^{4}}\left(3\left(4\beta^{2}\left(48\left(3\varepsilon+5\right)\beta^{2}-41\varepsilon+41\right)-153\varepsilon-7\right)\frac{\arctan\left(2\beta\right)}{2\beta}\right.\\
 & -\left.\frac{4\beta^{2}\left(16\left(9\varepsilon+31\right)\beta^{2}-429\varepsilon+109\right)-459\varepsilon-21}{\left(4\beta^{2}+1\right)}\right),\\
\varphi_{6}^{(a)} & =\frac{1}{3072\beta^{4}}\left(3\left(4\left(163\varepsilon-3\right)\beta^{2}-153\varepsilon-7\right)\frac{\arctan\left(2\beta\right)}{2\beta}\right.\\
 & -\left.\frac{4\beta^{2}\left(64\left(21\varepsilon+19\right)\beta^{2}+183\varepsilon-23\right)-459\varepsilon-21}{\left(4\beta^{2}+1\right)}\right),\\
\varphi_{7}^{(a)} & =-\frac{1}{48\beta^{4}}\left(3\left(2\beta^{2}\left(4\left(\varepsilon+1\right)\beta^{2}-3\varepsilon+3\right)-10\varepsilon+5\right)\frac{\arctan\left(2\beta\right)}{2\beta}\right.\\
 & -\left.\frac{\left(2\beta^{2}\left(20\left(\varepsilon+1\right)\beta^{2}-49\varepsilon+29\right)-30\varepsilon+15\right)}{\left(4\beta^{2}+1\right)}\right),\end{align*}
\begin{align*}
\varphi_{8}^{(a)} & =\frac{1}{48\beta^{4}}\left(3\left(2\varepsilon\left(2\beta^{2}-1\right)+1\right)\frac{\arctan\left(2\beta\right)}{2\beta}-\right.\\
 & -\left.\frac{\left(4\beta^{2}\left(8\left(\varepsilon+1\right)\beta^{2}-\varepsilon+2\right)-6\varepsilon+3\right)}{\left(4\beta^{2}+1\right)}\right),\\
\varphi_{9}^{(a)} & =\frac{1}{48\beta^{4}}\left(3\left(2\beta^{2}\left(4\left(\varepsilon+1\right)\beta^{2}-\varepsilon+1\right)-2\varepsilon+1\right)\frac{\arctan\left(2\beta\right)}{2\beta}-\left(2\left(\varepsilon+1\right)\beta^{2}-6\varepsilon+3\right)\right),\\
\varphi_{10}^{(a)} & =\frac{1}{48\beta^{4}}\left(\left(4\left(\varepsilon+1\right)\beta^{2}+6\varepsilon-3\right)-3\left(2\varepsilon\left(2\beta^{2}+1\right)-1\right)\frac{\arctan\left(2\beta\right)}{2\beta}\right).\end{align*}

The nonlinear turbulent diffusion of the LSMF in (\ref{eq:eta-quench}
) is expressed with help of the following functions

\begin{align*}
\varphi_{1} & =\frac{\left(\varepsilon-1\right)}{16\beta^{2}}\left(3\frac{\arctan\left(2\beta\right)}{\beta}-2\frac{\left(8\beta^{2}+3\right)}{\left(4\beta^{2}+1\right)}\right),\\
\varphi_{2} & =\frac{\left(\varepsilon+1\right)}{32\beta^{2}}\left(\left(4\beta^{2}+3\right)\frac{\arctan\left(2\beta\right)}{\beta}-3\right),\\
\varphi_{3} & =\frac{1}{8\beta^{2}}\left(\frac{\arctan\left(2\beta\right)}{\beta}-2\right).\end{align*}
The effect of slow rotation and nonuniform LSMF on the MEMF
 (eq. (\ref{eq:oxj:nl}))
is expressed with help of the following functions
 \begin{eqnarray*}
\varphi_{1}^{(w)} & = & -\frac{1}{6144\beta^{4}}\left(\frac{4\beta^{2}\left(4\beta^{2}\left(4\left(51\varepsilon-371\right)\beta^{2}+291\varepsilon+29\right)-219\varepsilon+1819\right)-315\varepsilon+1275)}{\left(4\beta^{2}+1\right)^{2}}\right.\\
 & + & \left.3\left(8\beta^{2}\left(2\left(17\varepsilon+47\right)\beta^{2}-51\varepsilon+51\right)+105\varepsilon-425\right)\frac{\arctan\left(2\beta\right)}{2\beta}\right),\\
\varphi_{2}^{(w)} & = & -\frac{1}{6144\beta^{4}}\left(3\left(8\beta^{2}\left(2\left(37\varepsilon+27\right)\beta^{2}+99\varepsilon-99\right)+105\varepsilon-425\right)\frac{\arctan\left(2\beta\right)}{2\beta}\right.\\
 & - & \left.\frac{4\beta^{2}\left(4\beta^{2}\left(4\left(273\varepsilon+47\right)\beta^{2}+1269\varepsilon-1589\right)+1119\varepsilon-2719\right)+315\varepsilon-1275}{\left(4\beta^{2}+1\right)^{2}}\right),\\
\varphi_{3}^{(w)} & = & \frac{1}{6144\beta^{4}}\left(3\left(24\beta^{2}\left(2\left(21\varepsilon+43\right)\beta^{2}+125\varepsilon-125\right)+735\varepsilon-2975\right)\frac{\arctan\left(2\beta\right)}{2\beta}\right.\\
 & - & \left.\frac{4\beta^{2}\left(4\beta^{2}\left(4\left(1347\varepsilon+125\right)\beta^{2}+5115\varepsilon-8123\right)+5925\varepsilon-17125\right)+2205\varepsilon-8925}{\left(4\beta^{2}+1\right)^{2}}\right),\\
\varphi_{4}^{(w)} & = & \frac{1}{6144\beta^{4}}\left(\frac{16\beta^{2}\left(\left(321\varepsilon-1\right)\beta^{2}+165\varepsilon-325\right)+315\varepsilon-1275}{\left(4\beta^{2}+1\right)}\right.\\
 & - & \left.3\left(8\beta^{2}\left(2\left(21\varepsilon+43\right)\beta^{2}+75\varepsilon-75\right)+105\varepsilon-425\right)\frac{\arctan\left(2\beta\right)}{2\beta}\right),\\
\varphi_{6}^{(w)} & = & \varphi_{5}^{(w)}=\varphi_{4}^{(w)},\\
\varphi_{7}^{(w)} & = & \frac{1}{6144\beta^{4}}\left(3\left(8\beta^{2}\left(2\left(37\varepsilon+27\right)\beta^{2}+33\varepsilon-33\right)+21\varepsilon-85\right)\frac{\arctan\left(2\beta\right)}{2\beta}\right.\\
 & - & \left.\frac{16\beta^{2}\left(\left(81\varepsilon-17\right)\beta^{2}+60\varepsilon-92\right)+63\varepsilon-255}{\left(4\beta^{2}+1\right)}\right),\\
\varphi_{8}^{(w)} & = & \frac{1}{6144\beta^{4}}\left(3\left(8\beta^{2}\left(2\left(17\varepsilon+47\right)\beta^{2}-17\varepsilon+17\right)+21\varepsilon-85\right)\frac{\arctan\left(2\beta\right)}{2\beta}\right.\\
 & + & \left.\frac{16\beta^{2}\left(\left(51\varepsilon-115\right)\beta^{2}+15\varepsilon+17\right)-63\varepsilon+255}{\left(4\beta^{2}+1\right)}\right).\end{eqnarray*}
\begin{eqnarray*}
  \varphi_1^{(s)} & = & \frac{1}{192 \beta^4 } \left( 3 \left( - 20
  \varepsilon + 3 \beta^2  \left( - \varepsilon + 4 \beta^2  \left(
  \varepsilon - 1 \right) - 7 \right) + 10 \right) \frac{\arctan \left( 2
  \beta \right)}{2 \beta} \right.\\
  & - & \left. \frac{\left( \beta^2  \left( 8 \beta^2  \left( 2 \beta^2 
  \left( 23 \varepsilon - 55 \right) - 67 \varepsilon - 25 \right) - 409
  \varepsilon + 137 \right) - 60 \varepsilon + 30 \right)}{\left( 4 \beta^2 +
  1 \right)^2} \right),\\
  \varphi_2^{(s)} & = & \frac{1}{192 \beta^4 } \left( \frac{\left( \beta^2 
  \left( 4 \beta^2  \left( 7 \varepsilon - 23 \right) - 35 \varepsilon - 5
  \right) - 12 \varepsilon + 6 \right)}{\left( 4 \beta^2 + 1 \right)}
  \right.\\
  & - & \left. 3 \left( - 4 \varepsilon + \beta^2  \left( - \varepsilon + 12
  \beta^2  \left( \varepsilon - 1 \right) - 7 \right) + 2 \right)
  \frac{\arctan \left( 2 \beta \right)}{2 \beta} \right),\\
  \varphi_3^{(s)} & = & \frac{\left( \varepsilon + 1 \right)}{16 \beta^2 }
  \left( \left( 4 \beta^2 + 3 \right)  \frac{\arctan \left( 2 \beta \right)}{2
  \beta} - 3 \right),\\
  \varphi_4^{(s)} & = & - \frac{\left( \varepsilon + 1 \right)}{8 \beta^2 }
  \left( \frac{\arctan \left( 2 \beta \right)}{2 \beta} - 1 \right),\\
  \varphi_5^{(s)} & = & \frac{1}{96 \beta^4 } \left( 8 \beta^2  \left(
  \varepsilon + 1 \right) - 6 \varepsilon + 3 - 3 \left( 4 \beta^2 + 1 - 2
  \varepsilon \right) \frac{\arctan \left( 2 \beta \right) }{2 \beta} \right)
\end{eqnarray*}

The quenching functions of the current helicity effects obtained in
the paper are
\begin{eqnarray*}
\varphi_{1}^{(h)} & = &
\frac{1}{4\beta^{2}}\left(1-\frac{\arctan\left(2\beta\right)}{2\beta}\right),\\
  \varphi_2^{(h)} & = & \frac{1}{64 \beta^2 } \left( 3 \left( 4 \beta^2 + 5
  \right)  \frac{\arctan \left( 2 \beta \right)}{2 \beta} - \frac{5 \left( 4
  \beta^2 + 3 \right)}{ \left( 4 \beta^2 + 1 \right)} \right),\\
  \varphi_3^{(h)} & = & - \frac{1}{192 \beta^4 } \left( 3 \left( 12 \beta^4 -
  \beta^2 - 8 \right) \frac{\arctan \left( 2 \beta \right)}{2 \beta} + \frac{
  \left( 4 \beta^4 + 67 \beta^2 + 24 \right)}{ \left( 4 \beta^2 + 1 \right)}
  \right),\\
  \varphi_4^{(h)} & = & \frac{1}{96 \beta^4 } \left( 3 \left( 12 \beta^4 - 3
  \beta^2 - 20 \right)  \frac{\arctan \left( 2 \beta \right)}{2 \beta} -
  \frac{\left( 368 \beta^6 - 536 \beta^4 - 409 \beta^2 - 60 \right)}{ \left( 4
  \beta^2 + 1 \right)^2} \right).
\end{eqnarray*}
%\varphi_{2}^{(h)} & = & \frac{1}{128\beta^{4}}\left(3\left(28\beta^{2}+13\right)\frac{\arctan\left(2\beta\right)}{2\beta}-\left(32\beta^{2}+39\right)\right),\\
%\varphi_{3}^{(h)} & = & \frac{1}{128\beta^{4}}\left(\frac{\left(64\beta^{4}-68\beta^{2}-39\right)}{\left(4\beta^{2}+1\right)}-3\left(64\beta^{4}+12\beta^{2}-13\right)\frac{\arctan\left(2\beta\right)}{2\beta}\right),\\
%\varphi_{4}^{(h)} & = & \frac{1}{128\beta^{4}}\left(3\left(64\beta^{4}-36\beta^{2}-65\right)\frac{\arctan\left(2\beta\right)}{2\beta}-\frac{\left(320\beta^{4}-628\beta^{2}-195\right)}{\left(4\beta^{2}+1\right)}\right).\end{eqnarray*}
The magnetic quenching functions for the shear-current effects are
\begin{eqnarray*}
  \varphi_1^{(V)} & = & \frac{1}{(8 \beta)^4} \left( \left( 16 \beta^2  \left(
  \beta^2  \left( 125 \varepsilon + 43 \right) - 25 \varepsilon - 3 \right) -
  \varepsilon - 87 \right) \frac{\arctan \left( 2 \beta \right)}{2 \beta}
  \right.\\
  & - & \left. \frac{\left( 8 \beta^2  \left( 2 \beta^2  \left( 128 \beta^2 
  \left( 3 \varepsilon + 13 \right) - 19 \varepsilon - 709 \right) - 755
  \varepsilon - 525 \right) - 15 \varepsilon - 1305 \right)}{15 \left( 4
  \beta^2 + 1 \right)} \right),\\
  \varphi_2^{(V)} & = & \frac{1}{2 (8 \beta)^4} \left( \left( 8 \beta^2 
  \left( 2 \beta^2  \left( 151 \varepsilon - 15 \right) - 15 \varepsilon - 57
  \right) - 125 \varepsilon + 165 \right) \frac{\arctan \left( 2 \beta
  \right)}{2 \beta}  \right.\\
  & - & \left. \frac{\left( 16 \beta^2  \left( \beta^2  \left( 443
  \varepsilon - 339 \right) - 85 \varepsilon - 3 \right) - 375 \varepsilon +
  495 \right)}{3 \left( 4 \beta^2 + 1 \right) } \right),\\
  \varphi_3^{(V)} & = & - \frac{1}{2 (8 \beta)^4} \left( \left( 8 \beta^2 
  \left( 2 \beta^2  \left( 247 \varepsilon - 367 \right) + 193 \varepsilon -
  393 \right) - 125 \varepsilon + 165 \right) \frac{\arctan \left( 2 \beta
  \right)}{2 \beta} \right.\\
  & + & \frac{\left( 16 \beta^2  \left( \beta^2  \left( 512 \beta^2  \left(
  \varepsilon + 3 \right) - 1563 \varepsilon + 2739 \right) - 227 \varepsilon
  + 507 \right) + 375 \varepsilon - 495 \right)}{3 \left( 4 \beta^2 + 1
  \right)} \left), \right.\\
  \varphi_4^{(V)} & = & \frac{1}{2 (4 \beta)^4} \left( \left( \beta^2  \left(
  4 \beta^2  \left( 17 \varepsilon - 5 \right) + 5 \varepsilon + 51 \right) +
  10 \varepsilon - 23 \right) \frac{\arctan \left( 2 \beta \right) }{2 \beta}
  \right.\\
  & + & \left. \frac{\left( \beta^2  \left( 4 \beta^2  \left( 64 \beta^2 
  \left( \varepsilon + 15 \right) - 609 \varepsilon - 235 \right) - 475
  \varepsilon + 155 \right) - 150 \varepsilon + 345 \right)}{15 \left( 4
  \beta^2 + 1 \right)} \right),\\
  \varphi_5^{(V)} & = & - \frac{1}{2 (4 \beta)^4} \left( \left( 8 \beta^2 
  \left( \left( 2 \beta^2 + 5 \right) \varepsilon - 6 \right) - 7 \varepsilon
  + 4 \right) \frac{\arctan \left( 2 \beta \right) }{2 \beta} \right.\\
  & + & \left. \frac{\left( 16 \beta^2  \left( \beta^2  \left( \left( 8 \beta
  - 7 \right)  \left( 8 \beta + 7 \right) \varepsilon + 80 \right) - 20
  \varepsilon + 35 \right) + 105 \varepsilon - 60 \right)}{15 \left( 4 \beta^2
  + 1 \right)} \right),
\end{eqnarray*}
\begin{eqnarray*}
  \varphi_6^{(V)} & = & - \frac{1}{2 (4 \beta)^4} \left( \left( 8 \beta^2 
  \left( 2 \beta^2  \left( 3 \varepsilon - 4 \right) + 7 \varepsilon + 10
  \right) + 35 \varepsilon - 40 \right) \frac{\arctan \left( 2 \beta \right)
  }{2 \beta} \right.\\
  & - & \left. \frac{ \left( 16 \beta^2  \left( \beta^2  \left( 64 \beta^2 
  \left( \varepsilon + 1 \right) + 23 \varepsilon + 44 \right) + 28
  \varepsilon - 5 \right) + 105 \varepsilon - 120 \right) }{3 \left( 4 \beta^2
  + 1 \right)} \right),\\
  \varphi_7^{(V)} & = & \frac{1}{(4 \beta)^2} \left( \left( 4 \beta^2  \left(
  8 \varepsilon - 7 \right) - 4 \varepsilon - 41 \right) \frac{\arctan \left(
  2 \beta \right) }{2 \beta} \right.\\
  & - & \left. \frac{ \left( 8 \beta^2  \left( 2 \beta^2  \left( 16 \beta^2 
  \left( 3 \varepsilon + 1 \right) + 48 \varepsilon - 67 \right) + 2
  \varepsilon - 113 \right) - 12 \varepsilon - 123 \right)}{3 \left( 4 \beta^2
  + 1 \right)^2} \right),\\
  \varphi_8^{(V)} & = & - \frac{1}{128 \beta^4} \left(  \left( 8 \beta^2 
  \left( 2 \beta^2  \left( 4 \varepsilon + 5 \right) + 3 \right) + 5 \right)
  \frac{\arctan \left( 2 \beta \right) }{2 \beta} - \frac{ \left( 16 \beta^2 
  \left( \beta^2  \left( 12 \varepsilon + 25 \right) + 7 \right) + 15
  \right)}{3 \left( 4 \beta^2 + 1 \right)} \right),\\
  \varphi_9^{(V)} & = & - \frac{1}{(8 \beta)^4} \left( \left( 8 \beta^2 
  \left( 2 \beta^2  \left( 9 \varepsilon + 47 \right) - 11 \varepsilon + 19
  \right) + 25 \varepsilon - 65 \right) \frac{\arctan \left( 2 \beta \right)
  }{2 \beta} \right.\\
  & - & \left. \frac{1}{15} \left( 4 \beta^2  \left( 256 \beta^2  \left(
  \varepsilon + 1 \right) - 455 \varepsilon + 895 \right) + 375 \varepsilon -
  975 \right) \right),\\
  \varphi_{10}^{(V)} & = & - \frac{1}{128 \beta^2} \left( \left( 4 \beta^2 
  \left( 3 \varepsilon - 11 \right) + 13 \varepsilon - 21 \right) 
  \frac{\arctan \left( 2 \beta \right) }{2 \beta} + \frac{1}{3} \left( 16
  \beta^2  \left( \varepsilon + 3 \right) - 39 \varepsilon + 63 \right)
  \right),\\
  \varphi_{11}^{(V)} & = & - \frac{\left( \varepsilon + 1 \right)}{48 \beta^2}
  \left( 3 \left( 4 \beta^2 + 1 \right)  \frac{\arctan \left( 2 \beta \right)
  }{2 \beta} - 8 \beta^2 - 3 \right) .
\end{eqnarray*}

The magnetic quenching functions for the current helicity evolution
equation:
\begin{eqnarray*}
  \psi_{1} & = & -\frac{\varphi^{(s)}_3}{\left(\varepsilon+1\right)},\\
  \psi_{2} & = &
  -\frac{1}{768\beta^{2}}\left(3\left(12\left(\varepsilon-1\right)\beta^{2}-21\varepsilon+5\right)
    \frac{\arctan\left(2\beta\right)}{2\beta} \right.\\
  &+&\left.
\frac{\left(4\beta^{2}\left(32\beta^{2}\left(\varepsilon+1\right)
+65\varepsilon-1\right)+63\varepsilon-15\right)}{\left(4\beta^{2}+1\right)}\right),\\
  \psi_{3} & = &  \frac{1}{48\beta^{2}}\left(3\left(\varepsilon-1\right)\frac{\arctan\left(2\beta\right)}{2\beta}
    -\left(4\beta^{2}\left(\varepsilon+1\right)+3\varepsilon-3\right)\right),\\ 
  \psi_4 & = & \frac{1}{192 \beta^4 } \left( \frac{\left. \left( 4 \beta^2 
  \left( 15 \varepsilon + 32 \beta^2 + 5 \right) + 9 \varepsilon + 3 \right)
  \right)}{ \left( 4 \beta^2 + 1 \right)} - 3 \left( 3 \varepsilon + 1 \right)
  \left( 4 \beta^2 + 1 \right) \frac{\arctan \left( 2 \beta \right)}{2 \beta}
  \right),\\
  \psi_5 & = & \frac{1}{8 \left( 2 \beta \right)^4 } \left( \left( 11
  \varepsilon + 4 \beta^2  \left( - \varepsilon + 8 \beta^2  \left(
  \varepsilon - 1 \right) + 5 \right) - 7 \right) \frac{\arctan \left( 2 \beta
  \right)}{2 \beta} \right.\\
  & - & \left. \frac{1}{15} \left( 8 \beta^2  \left( 8 \beta^2  \left(
  \varepsilon + 1 \right) - 35 \varepsilon + 55 \right) + 165 \varepsilon -
  105 \right) \right),\\
  \psi_6 & = & \frac{\left( 4 \beta^2 + 4 - 3 \varepsilon \right)}{48 \beta^4
  } \left( 3 \frac{\arctan \left( 2 \beta \right)}{2 \beta} - \frac{\left.
  \left( 8 \beta^2 + 3 \right) \right) }{\left( 4 \beta^2 + 1 \right)} \right)
\end{eqnarray*}

\subsection*{Appendix B. Comparison with some of results given in the paper by
R\"adler \& Stepanov (2006).}

This part of the article contains the comparison some of our results
with those from RS06. We apply the
mixing-length  (MLT) approximation to expressions obtained in RS06.
In this procedure we replace  the spectrum
of turbulent fields by the single-scaled function
of the form $\delta\left(k-\ell_{c}^{-1}\right)\delta\left(\omega\right)$,  where $\ell_{c}$ 
is the correlation length of the turbulence and 
put $\eta k^{2}=\nu k^{2}=\tau_{c}^{-1}$,\citep{kit:1991}.

\paragraph*{The effect of stratification and shear.}

The structure of the electromotive force obtained by RS06 can be reproduced
if we decompose the gradient of the large-scale flow $\overline{V}_{i,j}$
into symmetric and antisymmetric parts via
 \begin{equation}
\overline{V}_{i,j}=D_{ij}-\frac{1}{2}\varepsilon_{ijn}W_{n},\label{sh-dec}\end{equation}
where $W_{i}=\varepsilon_{inm}\overline{V}_{m,n}$ is the large-scale
vorticity and $D_{ij}=\left(\overline{V}_{i,j}+\overline{V}_{j,i}\right)/2$ is the rate
of strain tensor. After substitution (\ref{sh-dec}) to (\ref{emfas1})
we obtain 
 \begin{eqnarray}
\mathcal{E}_i^{(s)}&=&\left(\varepsilon_{inm} U_k \overline{B}_{n}
D_{mk} A_4 - \frac{A_4}{2}  \left( \mathbf{W \cdot \overline{B}}
\right) U_i + \left(\mathbf{U \cdot \overline{B}} \right)
\left( A_3 - \frac{A_2}{2} + \frac{A_1}{2} + \frac{A_4}{2} \right)
W_i\label{appB1}\right.\\
& +&\left.\varepsilon_{inm} \overline{B}_k D_{nk}U_m  \left( {A_2} + {A_1} \right) + \left(
\mathbf{W \cdot U} \right) \overline{B}_i  \left( \frac{A_2}{2} -
\frac{A_1}{2} \right)\right)\langle u^{(0)2}\rangle.\nonumber\\
&+&\tau_{c}^{2}\frac{h_{\mathcal{C}}^{\left(0\right)}}{6}
\left(\frac{5}{2}\left(\mathbf{W}\times\mathbf{\overline{B}}\right)_i
-\frac{23}{5}D_{ik}\overline{B}_k\right)\nonumber.
\end{eqnarray}
Using (\ref{appB1}) we find 
 \begin{eqnarray}
 \tau_{c}^{-2}\tilde{\alpha}_{1}^{(W)}&=&
 \frac{1}{2}\tau_{c}^{-2}\left(A_{1}-A_{2}\right)=0,\\ 
 \tau_{c}^{-2}\tilde{\alpha}_{2}^{(W)}&=&
 - \frac{1}{2}\tau_{c}^{-2}\left(A_3+ \frac{1}{2}\left(A_{1}-A_{2}\right)\right)=-\frac{1}{12},\\
 \tau_{c}^{-2}\tilde{\gamma}^{(W)}&=&
 - \frac{1}{2}\tau_{c}^{-2}\left(A_3+A_4+ \frac{1}{2}\left(A_{1}-A_{2}\right)\right)=0,\\
 \tau_{c}^{-2}\tilde{\gamma}^{(D)}&=&
  \frac{1}{2}\tau_{c}^{-2}\left(3A_4-A_{1}-A_{2}\right)=-\frac{11}{60},\\
\tau_{c}^{-2}\tilde{\alpha}^{(D)}&=&
 \frac{1}{2}\tau_{c}^{-2}\left(A_4-A_{1}-A_{2}\right)=-\frac{1}{60},\\ 
\end{eqnarray}
where we put $\varepsilon=0$.
After applying the MLT to results obtained in RS06 we find $\tau_{c}^{-2}
  \widetilde{\alpha}_{1}^{(W)}=19/120$, 
 $\tau_{c}^{-2}\widetilde{\alpha}_{2}^{(W)}=-7/240$,
 $\tau_{c}^{-2}\widetilde{\gamma}^{(W)}=-1/48$, $\tau_{c}^{-2}\widetilde{\gamma}^{(D)}=-39/120$, $\tau_{c}^{-2}\widetilde{\alpha}^{(D)}=-21/120$.

\paragraph*{The effect of nonuniform LSMF and shear.}

For the shear-current effect, after substitution of (\ref{sh-dec})
to (\ref{difs1}) we arrive to the following representation of $\mathcal{E}_{i}^{(V)}$,

\begin{align}
\mathcal{E}_{i}^{(V)} & =\left\{ \frac{C_{3}-C_{4}}{2}\left(\mathbf{W}\cdot\nabla\right)\overline{B}_{i}+\frac{C_{1}-C_{2}}{2}\nabla_{i}\left(\mathbf{W\cdot}\overline{\mathbf{B}}\right)\right\} \left\langle u^{(0)2}\right\rangle \label{difs2}\\
 & +\varepsilon_{inm}\left\{
   \left(C_{1}+C_{2}\right)\overline{B}_{n,l}
+\left(C_{3}+C_{4}\right)\overline{B}_{l,n}\right\} D_{ml}\left\langle u^{(0)2}\right\rangle .\nonumber \end{align}
Using this formula we obtain 
\begin{eqnarray}
\tau_{c}^{-2}\tilde{\delta}^{(W)}&=& 
\frac{1}{4}\tau_{c}^{-2}\left(C_{3}-C_{4}-C_{1}+C_{2}\right)=\frac{1}{12},\\
\tau_{c}^{-2}\tilde{\kappa}^{(W)}&=& 
\frac{1}{2}\tau_{c}^{-2}\left(C_{4}+C_{2}-C_{1}-C_{3}\right)=-\frac{4}{15},\\  
\tau_{c}^{-2}\tilde{\kappa}^{(D)}&=&
- \frac{1}{2}\tau_{c}^{-2}\left(C_{1}+C_{2}+C_{3}+C_{4}\right)=\frac{3}{10},
\\ 
 \tau_{c}^{-2}\tilde{\beta}^{(D)}&=&
- \frac{1}{2}\tau_{c}^{-2}\left(C_{1}+C_{2}-C_{3}-C_{4}\right)=0,
\end{eqnarray}
where we put  $\varepsilon=0$ in $C_{1-4}$. After applying the MLT to
results in RS06 we find
$\tau_{c}^{-2}\tilde{\delta}^{(W)}=1/12$,
$\tau_{c}^{-2}\tilde{\kappa}^{(W)}=-1/30$,
$\tau_{c}^{-2}\tilde{\kappa}^{(D)} = 13/30$ and
$\tau_{c}^{-2}\tilde{\beta}^{(D)} = 7/60$.

\end{document}